\documentclass[fleqn,usenatbib]{mnras}

\usepackage{newtxtext,newtxmath}

\usepackage[T1]{fontenc}

\DeclareRobustCommand{\VAN}[3]{#2}
\let\VANthebibliography\thebibliography
\def\thebibliography{\DeclareRobustCommand{\VAN}[3]{##3}\VANthebibliography}

\usepackage{graphicx}	% Including figure files
\usepackage{amsmath}	% Advanced maths commands
\newcommand{\angstrom}{\textup{\AA}}
\usepackage{siunitx}
\usepackage{booktabs}
\usepackage{gensymb}
\usepackage[para]{threeparttable}
\usepackage[table,xcdraw]{xcolor}
\title[Revisiting the star cluster Fornax~6]{Revisiting the enigmatic sixth star cluster in the Fornax dwarf spheroidal galaxy}

\author[C. Crociati et al.]
{C. Crociati,$^{1}$\thanks{E-mail: ccrociat@ed.ac.uk}
A. M. N.~Ferguson,$^{1}$
R. Pascale,$^{2}$
S. Kamann,$^{3}$
S. Martocchia,$^{4}$
J. M. Howell,$^{1}$
P. Kuzma,$^{1}$ \newauthor
D. Mackey,$^{5}$
S. Okamoto,$^{6,7}$
and M. Onodera$^{6,7}$
\\
$^{1}$Institute for Astronomy, University of Edinburgh, Royal Observatory, Blackford Hill, Edinburgh EH9 3HJ, UK  \\
$^{2}$INAF – Osservatorio di Astrofisica e Scienza dello Spazio di Bologna, Via Gobetti 93/3, 40129 Bologna, Italy \\
$^{3}$Astrophysics Research Institute, Liverpool John Moores University, IC2 Liverpool Science Park, 146 Brownlow Hill, Liverpool L3 5RF, UK \\
$^{4}$Aix Marseille Université, CNRS, CNES, LAM, Marseille, France \\
$^{5}$Independent Researcher, Charnwood, Canberra, ACT 2615, Australia\\
$^{6}$Subaru Telescope, National Astronomical Observatory of Japan, 2-21-1 Osawa, Mitaka, Tokyo 181-8588, Japan\\
$^{7}$Graduate Institute for Advanced Studies, SOKENDAI, 2-21-1 Osawa, Mitaka, Tokyo 181-8588, Japan}

\date{Accepted 2026 November 11. Received 2026 November 7; in original form 2026 August 27}

\pubyear{\the\year{}}

\newcommand{\ncp}{n_{\rm comp}}
\newcommand{\wi} {w_i}
\newcommand{\wii} {w_1}

\newcommand{\Sigmai}{\Sigma_i}

\newcommand{\SigmaFnx}{\Sigma_{\rm F6}}
\newcommand{\Rh}{R_h}

\newcommand{\xc}{x_{\rm 0}}
\newcommand{\yc}{y_{\rm 0}}

\newcommand{\Lprob}{\mathscr{L}}
\newcommand{\SSS}{\mathcal{S}}
\newcommand{\DD}{\mathcal{D}}
\newcommand{\EE}{\mathcal{E}}
\newcommand{\MM}{\mathcal{M}}
\newcommand{\VV}{\mathcal{V}}
\newcommand{\Area}{\mathcal{A}}
\newcommand{\LL}{\mathcal{L}}
\newcommand{\PP}{\mathcal{P}}
\newcommand{\PPfnx}{\mathcal{P}_{\rm F6}}
\newcommand{\PPBG}{\mathcal{P}_{\rm BG}}

\newcommand{\Rmax}{R_{\rm max}}

\newcommand{\dd}{{\rm d}}
\newcommand{\alphac}{\alpha_c}
\newcommand{\deltac}{\delta_c}

\newcommand{\xj}{x_j}
\newcommand{\yj}{y_j}
\newcommand{\vlosj}{v_{{\rm los},j}}
\newcommand{\vlos}{v_{\rm los}}
\newcommand{\metj}{{\rm [Fe/H]}_j}
\newcommand{\met}{{\rm [Fe/H]}}
\newcommand{\fnx}{{\rm F6}}

\newcommand{\BG}{{\rm BG}}
\newcommand{\kms}{\rm km \,s^{-1}}

\definecolor{lightgray}{gray}{0.9} 

\begin{document}
\label{firstpage}
\pagerange{\pageref{firstpage}--\pageref{lastpage}}
\maketitle

\begin{abstract}
%The abstract should briefly describe the aims, methods, and main results of the paper.
%It should be a single paragraph not more than 250 words (200 words for Letters).
%No references should appear in the abstract.
The Fornax dwarf spheroidal (dSph) galaxy is one of only two Local Group dSphs that host a population of globular star clusters (GCs), the present-day properties of which have often been used to investigate the nature of the dark matter. An additional overdensity of stars called Fornax 6, lying at a projected distance of $\sim$0.3 kpc from the galaxy centre, was recently identified as a sixth GC residing in Fornax. However, this conclusion was based on relatively shallow photometric observations and a limited number of spectra. Here, we reinvestigate the nature of Fornax 6 by analysing MUSE/Wide-Field-Mode observations alongside deep GMOS-S imagery. Using a sample of 131 spectra of red giant and red clump stars, we confirm the Fornax 6 overdensity as a distinct chemo-dynamical component with respect to the surrounding field population. Specifically, we identify 43 likely members associated with the cluster, from which we measure $\met=-0.63\pm0.03\,$dex, $\overline{\vlos} = 50.8^{+0.8}_{-0.7}\,\kms$, and $\sigma_v = 3.3^{+1.4}_{-1.9}\,\kms$. The main-sequence turn-off, observed here for the first time, strongly suggests that these stars are coeval and well described by a $\sim$3 Gyr old isochrone. The cluster is characterised by an irregular morphology, a large half-light radius ($\Rh=8.8^{+1.1}_{-1.3}\,$pc), a small flattening ($e<0.26$, 68\% upper limit), and low luminosity  ($M_V = -5.0 \pm 0.4$). Our improved characterisation of Fornax 6 supports its classification as a genuine low-mass cluster likely undergoing tidal disruption, making it the youngest and most metal-rich member of Fornax’s unique GC system.

\end{abstract}


%%%%%%%%%%%%%%%%%%%%%%%%%%%%%%%%%%%%%%%%%%%%%%%%%%

%%%%%%%%%%%%%%%%% BODY OF PAPER %%%%%%%%%%%%%%%%%%

\section{Introduction}
\label{sec: introduction}
Situated at a distance of $\sim$143 kpc \citep{McConnachie2012,oakes+2022}, the Fornax dwarf spheroidal (dSph) galaxy has long been recognised as a very intriguing Milky Way satellite.
Several photometric and spectroscopic investigations showed the coexistence of two main stellar populations (see, e.g., \citealt{stetson+1998, battaglia+2006, walker+2006,walker_penarrubia2011,delPino+2015,wang_fornax+2019}): a dynamically hot, spatially extended, metal-poor ($-2.5\lesssim \met \lesssim -1.3\,$dex) population which is traced by old stars ($t>10\,$Gyr) and a kinematically cold, centrally concentrated, and more metal-rich ($-1.2 \lesssim \met \lesssim -0.5\,$dex) population. This latter component includes both intermediate and very young (8 Gyr $\gtrsim t \gtrsim$ 200 Myr) stars and represents the bulk of the stellar population of Fornax. Moreover, chemo-dynamical evidence suggests that this metal-rich population can be split into two distinct subcomponents, peaking at $\met= -0.9\,$dex and $\met = -0.65\,$dex \citep{amorisco&evans2012}. The latter, more metal-rich subpopulation is found to be even kinematically colder and more centrally concentrated than the metal-intermediate one. Evidence of a misalignment between the angular momenta of different populations in Fornax suggests a past merger (\citealt{amorisco&evans2012, delPino+2017}), making it an interesting case for investigating hierarchical assembly on small scales.   

The detailed star formation history (SFH) of the central regions of Fornax has been investigated by \cite{rusakov+2021} using deep {\it Hubble} Space Telescope (HST) colour-magnitude diagrams (CMD).  They find evidence for three main star formation events. The first is a prolonged phase of star formation that began $\sim$12 Gyr ago and lasted until $\sim$8 Gyr. This is followed by a prominent burst of star formation $\sim$5 Gyr ago, and finally, a sequence of less intense episodes occurring at more recent times ($t < 3\,$Gyr). In their SFH solution, the three dominant star formation events are associated with the distinct metallicities observed for the three chemo-dynamical subpopulations.

Interestingly, along with the disrupting Sagittarius dwarf, Fornax is one of only two Local Group dSphs known to possess a system of globular clusters (GCs). It hosts five massive GCs, all of which have been studied in some detail with Hubble Space Telescope (HST) imagery (e.g., \citealt{buonanno+1998,buonanno+1999,mackey_gilmore2004,larsen+2014,deBoer_fraser2016,martocchia+2020,sarajedini2024}) as well as ground-based spectroscopy (e.g., \citealt{letarte+2006,larsen+2014,larsen+2022}). Four of the GCs (Fornax~1, 2, 3, and 5) appear to be coeval, to within $\sim$1 Gyr, with the oldest Galactic GCs and are amongst the most metal-poor GCs known in the local Universe, with $-2.5\leq \met \leq -2.0\,$dex (see \citealt{beasley+2019}). The fifth cluster (Fornax~4) may be $2-3\,$Gyr younger and is also considerably more metal-rich with $\met = -1.4\,$dex. These GCs are extremely informative as they can be used to signpost when the dwarf underwent significant periods of star formation, to pinpoint merger events \citep{leung+2020}, and to address how fundamental scaling relations - such as that between host galaxy halo mass and GCs system mass - behave in the very low mass regime (e.g., \citealt{forbes+2018}). 

Furthermore, reconciling the existence of these GCs with the dark-matter halo density profiles predicted by the cold dark-matter cosmological framework ($\Lambda$CDM) has long been considered a key test of the standard cosmological model. Since predicted dark-matter haloes have cuspy inner profiles \citep{navarro+1996,navarro+1997}, GCs are expected to be heavily affected by dynamical friction as they move through the dark-matter dominated Fornax ($M/L\sim 10$; \citealt{Kowalczyk+2019}). This process should cause orbital decay on a time-scale much shorter than the ages determined from their CMDs (e.g., \citealt{goerdt+2006,arca-sedda_capuzzo-dolcetta2016}). It is therefore surprising that these five GCs have not merged to form a nucleus at the centre of Fornax, but rather have survived a Hubble time to be observed today at projected radial distances of $0.1 - 1.1\,$kpc. 
Although a cuspy dark-matter halo model, under very specific conditions, can be reconciled with the long-lived system of GCs observed in Fornax (see, for example, \citealt{boldrini+2019, meadows+2020,shao+2021}), a standard solution to this dilemma is to assume a constant density profile at the centre (e.g., \citealt{cole+2012}), as independently shown by various studies fitting the dSph internal kinematics and structure (e.g., \citealt{walker_penarrubia2011, pascale+2018}).
%using equilibrium dynamical models that fit the dSph internal kinematics and structure (e.g., \citealt{pascale+2018}). 
More generally, this discrepancy with the CDM prediction, known as the cusp/core problem, is evident in other kinematical studies of dwarf galaxies (see \citealt{battaglia_nipoti2022} for a recent review).

Clearly, the present-day distribution and properties (age, metallicity, and mass) of Fornax's GCs are of considerable importance.  Recent work has revitalised a longstanding debate about the existence of an additional stellar cluster – dubbed Fornax~6 – lying at a projected distance of only $\sim$270 pc from the centre of Fornax. First described by \cite{shapley1939} as a “very faint cluster”, Fornax~6 has been subsequently referred to as a statistical field overdensity \citep{verner+1981,demers+1994} and a distant compact group of galaxies \citep{stetson+1998}. It disappeared from the literature until the study of \cite{wang+2019}, where they used photometry from DECam ($i\sim22.5$) to argue for a statistically-significant overdensity of red giant branch (RGB) stars.  However, their derived CMD is relatively shallow, reaching to just below the Fornax dSph red clump (RC), and exhibits a very large spread in colour, unlike that expected for a single stellar population. This suggests non-negligible contamination from field stars and/or background galaxies in their sample, and/or poor photometric accuracy. The Fornax~6 overdensity was also reported later using Gaia Data Release 2 data \citep{huang&koposov2021}.

In a follow-up study to \cite{wang+2019}, \cite{pace+2021} present Magellan spectroscopy of RGB stars in the vicinity of Fornax~6 and in the wider Fornax dSph. They adopted a mixture-model to assign membership of stars to Fornax~6 and identified 10-15 stars as probable members, although only 6 of those lie within the half-light radius measured by \cite{wang+2019}. They found that Fornax~6 has the same radial velocity as the Fornax dSph, confirming an association, but that it has a much higher velocity dispersion than would be expected for a similar GC. As a consequence, the derived mass-to-light ratio is anomalously high for a star cluster ($15<M/L_V<258$). Given the large ellipticity ($e=0.41\pm0.10$) measured by \cite{wang+2019}, \cite{pace+2021} concluded that the cluster is experiencing ongoing tidal disruption, therefore attributing the measured velocity dispersion to a non-equilibrium configuration. However, two other explanations for the high velocity dispersion exist. One is that the \cite{pace+2021} sample is contaminated by Fornax field interlopers, which is plausible given the low number of member stars they identify inside the half-light radius. Another more tantalising possibility is that Fornax~6 is made of field stars temporarily captured by a dark
substructure orbiting in the Fornax dSph potential \citep{penarrubia+2024}. Both of these hypotheses are supported by the high metallicity ($\met = -0.71\,$dex) and modest age ($\sim 2\,$Gyr) for Fornax~6 derived by \cite{pace+2021}, which are completely consistent with the surrounding field population (e.g., \citealt{deBoer+2012}). 

In this work, we revisit Fornax~6 with vastly improved data to better characterise the properties and nature of this putative star cluster. In particular, we leverage the statistical power of the integral-field spectrograph MUSE on the European Southern Observatory Very Large Telescope (ESO-VLT) along with deep, high spatial resolution imagery from Gemini/GMOS-S. The paper is organised as follows: in Section~\ref{sec: observation_reduction}, we describe our data sets and the reduction steps that were performed to derive our initial catalogues; in Section~\ref{sec: data_analysis}, we present a description of the final working catalogues; in Section~\ref{sec:charac}, we characterise the properties of the Fornax~6 overdensity, including its luminosity, morphology, metallicity and kinematics; in Section~\ref{sec: discussion}, we discuss our results in the context of existing literature and the Fornax dSph; and in Section~\ref{sec: summary_conclusions}, we summarise our main findings and draw our conclusions.

\begin{figure*}
    \centering
    \includegraphics[width=1\textwidth]{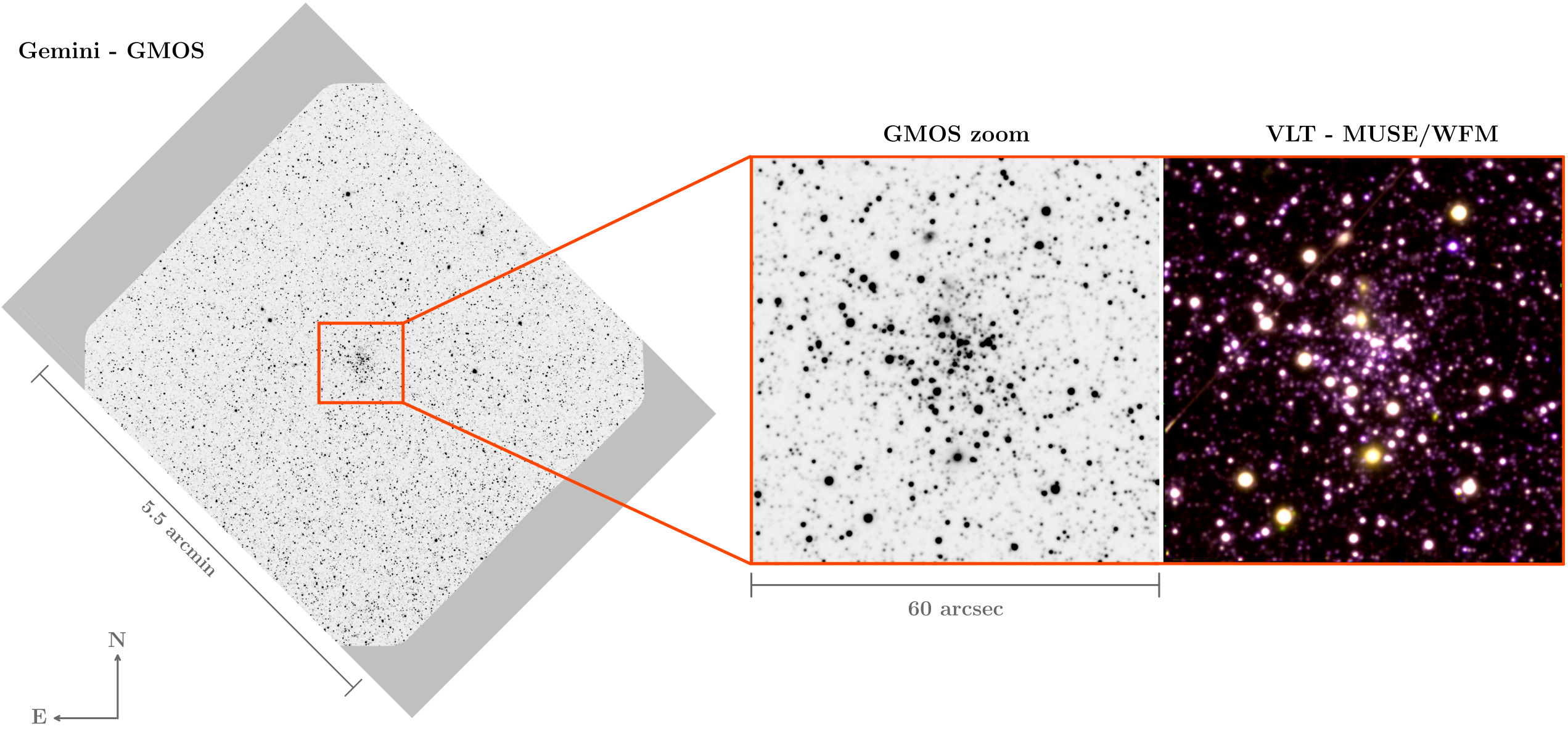}
    \caption{Stacked GMOS-S image in the $g$ filter centred on the sixth stellar cluster candidate of Fornax (Fornax~6). The inset shows the comparison between a zoom-in of the central $\sim 1\arcmin \times 1\arcmin$ area surrounding the overdensity and the MUSE false colour $gri$ image created by convolving the final MUSE datacube with SDSS filter response curves. North is up, east is to the left.}
\label{fig: FOV}
\end{figure*}

\section{Observations and data reduction}
\label{sec: observation_reduction}

\begin{figure*}
    \centering
    \includegraphics[width=0.8\textwidth]{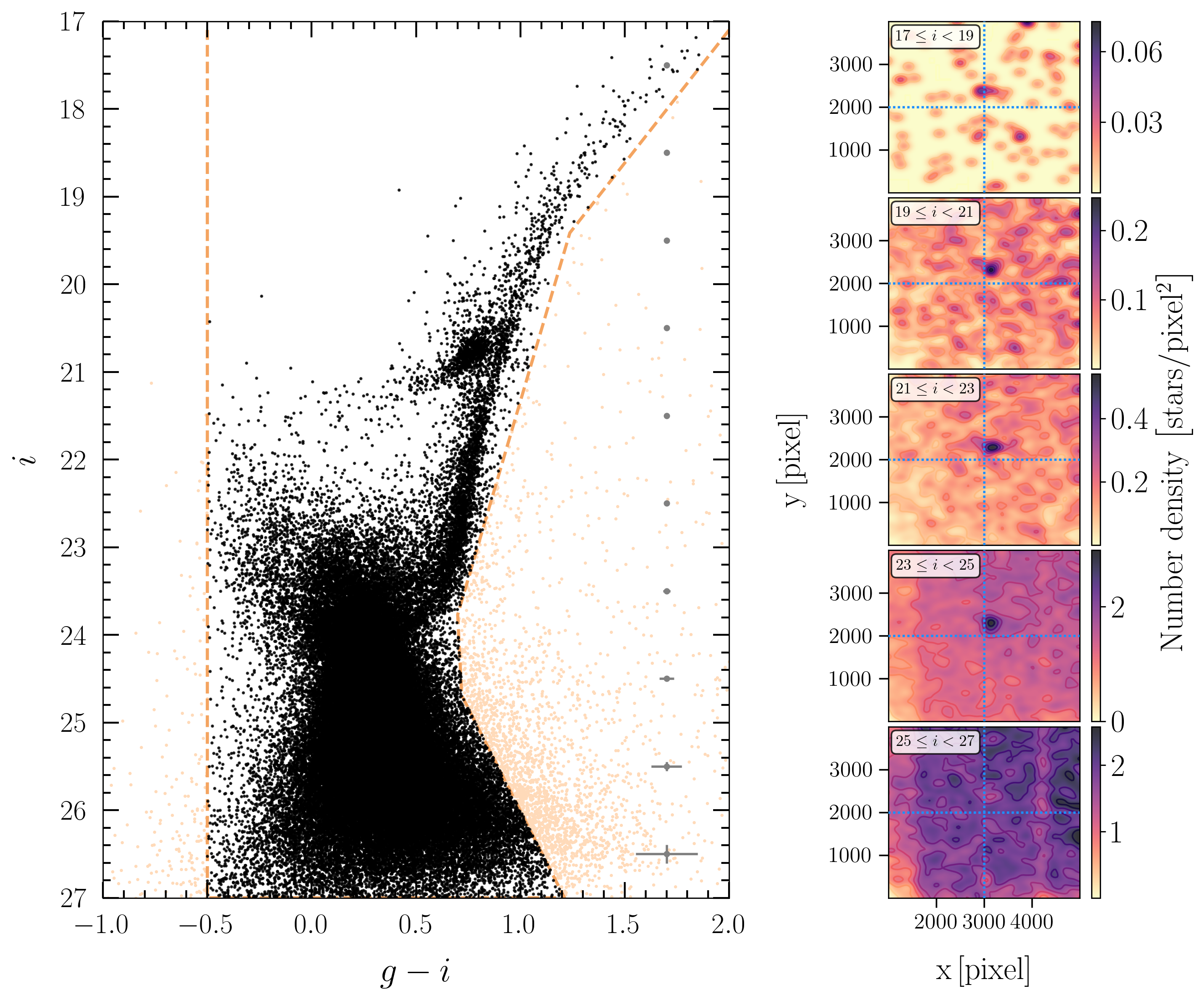}
    \caption{Left panel: GMOS $(i, g-i)$ CMD of stars within the total field-of-view after \emph{chi} and \emph{sharpness} cuts described in Section~\ref{sec: photometric_catalogue} have been applied. Mean colour and magnitude errors as derived from \texttt{DAOPHOT} are shown as grey error bars. Orange dots represent sources that were excluded from the analysis following the colour selection described in Section~\ref{sec: photometric_catalogue}. Right column: density maps of the final photometric catalogue (see black dots in the left panel) divided into 5 bins of 2 mag each, starting at $i=17$. The field-of-view was binned into $150 \times 150$ spatial bins, and the resulting 2D histogram was smoothed adopting a Gaussian kernel with a dispersion of $25\,$pixel. Horizontal dotted lines mark the centre of the density plots.}
\label{fig: cmd_density}
\end{figure*}

\subsection{MUSE data set}
\label{sec: muse_dataset}
The primary focus of this work is the analysis of spectroscopic data acquired with the ESO-VLT integral field spectrograph MUSE \citep{bacon+2010} in the wide-field mode (WFM) configuration. Specifically, this configuration translates into a field-of-view (FOV) of $1\arcmin\times 1\arcmin$ and a spatial sampling of $0.2\arcsec\,\text{pixel}^{-1}$. The wavelength range covered by the nominal filter of MUSE corresponds to $4800 - 9300\,\angstrom$, with a spectral sampling of $1.25\,\rm\angstrom$ and a resolution of $R\sim3100$ at $8000\,\angstrom$. The observations belong to the programme ID 0102.B-0911 (PI: Ferguson), and they were performed with the Ground Layer Adaptive Optics mode (i.e., WFM-AO) of the VLT Adaptive Optics Facility (AOF; \citealt{arsenault+2008}) through the GALACSI AO module \citep{stroebele+2012}. A total of $15\times1380$s exposures centred on Fornax~6 were acquired from November 2018 to September 2019, with a DIMM seeing ranging from $0.4\arcsec$ to $1.4\arcsec$. The raw exposures were reduced using the standard ESO pipeline \citep{weilbacher+2020}. For each of the 24 identical Integral Field Units (IFUs), we performed bias subtraction, flat-fielding, wavelength calibration, sky subtraction, astrometric and flux calibration, and heliocentric velocity correction. The single processed exposures were then combined together, creating the final datacube. The rightmost panel of Figure~\ref{fig: FOV} shows an image of the MUSE FOV created by convolving the final datacube with SDSS-$gri$ filter transmission curves. 

We extracted individual source spectra from the MUSE datacube using the software \texttt{PampelMuse} \citep{kamann+2013}. By exploiting the object coordinates from an input astrometric reference catalogue, this software can perform source deblending using a wavelength-dependent point spread function (PSF) model. We used the photometric catalogue derived from deep GMOS-S stacked images (see the following section) as our reference catalogue.

\subsection{GMOS-S data set}
\label{sec: reduction_GMOS}

We complemented the MUSE data with deep archival images obtained with the GMOS-S (hereafter, GMOS) instrument \citep{hook+2004} mounted at the $8\,$m Gemini South Telescope, acquired under the program GS-2007B-Q-23 (PI: Mackey). At the time of these observations (November 2007), the GMOS detector was composed of three EEV CCDs made of $2048\times4176\,$pixels, characterised by a pixel scale of $0.073\arcsec$ and a total FOV of $5.5\arcmin \times 5.5\arcmin$.
The data set is composed of nine exposures in the $i$ filter with $t_{exp}=240\,$s and four $g$ band exposures with $t_{exp}=660\,$s centred on the Fornax~6 overdensity (see Figure~\ref{fig: FOV}). The observations were performed under exquisite $0.5\arcsec$ seeing conditions and with a dither pattern of a few arcseconds, which partially covered the intra-chip gaps of $2.8\arcsec$. The full FOV in the $g$ band is shown in the left panel of Figure~\ref{fig: FOV}, together with a zoom-in of a $\sim60\arcsec \times 60\arcsec$ area surrounding the overdensity.

We performed PSF fitting on the detrended and stacked images retrieved from the Gemini archive. The photometric reduction was carried out by means of \texttt{DAOPHOT II} \citep{stetson1987}. First, we modelled the PSF of each $g$ and $i$ co-added image by using the \texttt{DAOPHOT/PSF} routine. We selected more than 200 bright, isolated, and unsaturated stars across the FOV and allowed the PSF to vary within the detector according to a quadratic polynomial spatial model. We applied the resulting PSF models to all the detected sources, defined as flux peaks above at least $3\sigma$ with respect to the local background, by using $\texttt{DAOPHOT/ALLSTAR}$.
Second, we combined these two band catalogues of instrumental magnitudes and positions using the cross-matching subroutines \texttt{DAOMATCH} and \texttt{DAOMASTER} (\citealt{stetson1993}), generating an input master list for \texttt{ALLFRAME} \citep{stetson1994} which provided the final photometry adopted in this work. %In this way, we ensured to maximise the information coming from both images, given that the \texttt{ALLFRAME} routine forces a fit at the corresponding positions of stars in the master list. 

We converted instrumental magnitudes into the DECam broad-band filters system by cross-matching our final catalogue with the publicly available Dark Energy Survey Data Release 2 (DES DR2; \citealt{abbott+2021}). We considered the sources classified as stars in the DES catalogue following the \texttt{spread model} photometric parameter. To minimise the effect of photometric errors and blending, we considered only stars brighter than $i_{\rm DES}=22$, for a total of $\sim 900$ common sources. Since no trend with colour was observed, we assumed as calibration zero-points the median of the distributions $g_{\rm DES} - g_{\rm GMOS}$ and $i_{\rm DES} - i_{\rm GMOS}$, after having removed clear outliers by performing an iterative $3\sigma$-clipping procedure. The rms of the best-fit relation is 0.09 and 0.05 in
$g$ and $i$, respectively.  Instrumental coordinates were transformed to absolute sky coordinates $(\alpha,\delta)$ via a cross-match with Gaia Data Release 3  \citep{gaia_mission,gaiadr3} through the software \texttt{CataXcorr} (P.~Montegriffo; software publicly available online\footnote{http://davide2.bo.astro.it/paolo/Main/CataPack.html}), which is designed to perform accurate astrometric solutions. Within radii of $R<6\arcmin$ from the centre of the central GMOS detector, we matched $\sim 1500$ sources. The resulting rms scatter of the solution is $<0.05\arcsec$ in both directions.

\section{Catalogues} 
\label{sec: data_analysis}

\subsection{Photometric catalogue and colour-magnitude diagram}
\label{sec: photometric_catalogue}

As a first step, we cleaned the photometric catalogue derived in Section~\ref{sec: reduction_GMOS} by rejecting sources on the basis of the $chi$ and $sharpness$ parameter distributions, which are parameters computed by \texttt{DAOPHOT} that quantify the quality of the PSF-fitting process. 
We divided the sample into 1 magnitude bins and, for each interval, retained sources within $3\sigma$ of the median value.
This procedure ensures a first-order cleaning of extended background galaxies, spurious artefacts, bad pixels, and cosmic rays. The resulting CMD is shown in the left panel of Figure~\ref{fig: cmd_density}. In addition, we excluded from the catalogue sources characterised by extreme red or blue colours following the orange dashed lines displayed in the CMD of Figure~\ref{fig: cmd_density}.  Visual inspection of the cleaned catalogue overlaid on the GMOS images confirmed that all obvious non-stellar sources had been excised from the catalogue.

The final photometric catalogue is represented by the black dots in Figure~\ref{fig: cmd_density}. This CMD is 4 magnitudes deeper than the one presented in \cite{wang+2019} based on DECam observations, and is comparable to deep literature CMDs obtained with HST \citep{rusakov+2021} and ESO-VLT/FORS1 \citep{delPino+2013}. It clearly shows the presence of old and intermediate MSTOs coupled with a blue plume of young stars. 

The right-hand panel of Figure~\ref{fig: cmd_density} shows the density distribution of the final catalogue at different $i$-band magnitude levels (from $i=$27 to 17, in steps of 2 magnitudes). These maps correspond to 2D histograms created by dividing the FOV into $33.5\times 27\,\rm pixel^2$, for a total of 150 bins in each direction. The maps were subsequently smoothed through the \texttt{scipy} Gaussian filter \citep{scipy}, adopting a Gaussian kernel of $\sigma=25\,\rm pixel$. In the same panels, we marked the centre of the density plots with blue dotted lines to guide the eye. A persistent overdensity emerges in the centre of the GMOS chip when looking at the magnitude range $25\leq i \leq 19$. 

\subsection{Spectroscopic catalogue}
\label{sec: spectroscopy}

\begin{figure}
    \centering
    \includegraphics[width=1\columnwidth]{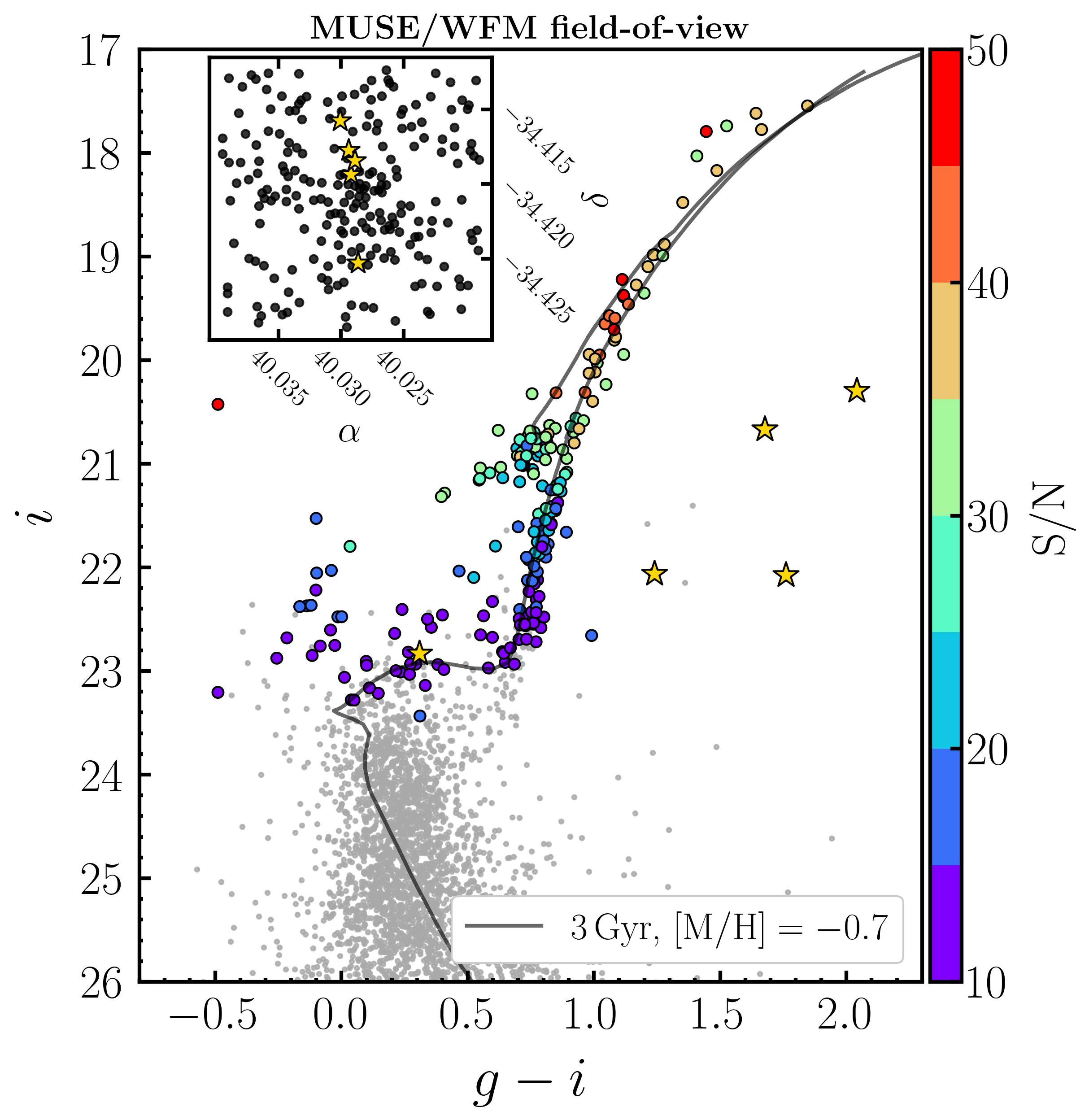}
    \caption{CMD of extracted sources colour-coded according to the average S/N calculated by \texttt{PampelMuse}. Grey points represent the final GMOS photometric catalogue presented in Section~\ref{sec: photometric_catalogue} within the FOV covered by MUSE. Yellow stars indicate objects with galaxy-like spectra. The inset shows the spatial distribution (north up, east to the left) of the extracted sources, where galaxy-like objects are marked as yellow stars.}
\label{fig: pampelmuse_results}
\end{figure}

The procedure described in Section~\ref{sec: muse_dataset} yielded the extraction of 224 spectra with signal-to-noise ($S/N$) higher than 10. Sources for which we recovered the spectrum are displayed in the GMOS CMD of Figure~\ref{fig: pampelmuse_results} as green dots colour-coded by the S/N measured by \texttt{PampelMuse}, together with all the stars photometered within the MUSE footprint (grey points). In addition, we show the spatial distribution of the extracted sources in the inset of the same plot, marked in black. 
Visual inspection of this pool of spectra confirmed the presence of a few background galaxies at the corresponding position of the overdensity, which are also visible by eye as extended sources and diffuse yellow objects in the zoom-in panels of Figure~\ref{fig: FOV}. Sources with galaxy-like spectra\footnote{Specifically, the identified galaxy spectra showed a prominent CaII H\&K absorption and [OII] emission lines.} are shown as yellow stars in Figure~\ref{fig: pampelmuse_results}, and they will be excluded from the spectral analysis. A more detailed discussion of these sources can be found in Section~\ref{sec: background_galaxies}.

To measure the line-of-sight velocity ($v_{\rm los}$) and the metallicity of the extracted stellar spectra, we employed the full-spectrum fitting code \texttt{SPEXXY} \citep{husser+2016}. This software is tailored to analyse MUSE spectra and has been successfully employed in many studies of resolved stellar populations obtained with this instrument (see, e.g., \citealt{kamann+2018}). By comparing the observed spectrum with a library of stellar models, the code determines the star's atmospheric parameters, including [Fe/H] and $v_{\rm los}$. Given the low resolution of MUSE, an initial guess for the effective temperature $T_{\rm eff}$ and the surface gravity log$\,g$ is highly recommended (see \citealt{husser+2016}). To this purpose, we considered a PARSEC isochrone \citep{bressan+2012} with $t=3\,$Gyr and $\rm [M/H] = -0.7\,dex$, which, by adopting one of the most recent Fornax distance modulus determinations ($\mu = 20.77$; \citealt{oakes+2022}) and a reddening value of $E(B-V) = 0.06$ in agreement with literature estimates (e.g., \citealt{martocchia+2020}), nicely reproduces the RGB and RC of the main stellar population (see the solid line in Fig.~\ref{fig: pampelmuse_results}). Initial guesses on the atmospheric parameters are then determined by projecting the observed stars onto the isochrone. Various stellar populations, with different ages and metallicities (see Sect.~\ref{sec: introduction}), are apparent on the CMD shown in Figure~\ref{fig: pampelmuse_results}, and \texttt{SPEXXY} is free to readjust the input atmospheric parameters as required. However, if the analysis is restricted to RGB and RC stars, the $T_{\rm eff}$ and log$\,g$ retrieved from the different isochrones are very similar. For the library of templates, we used the collection of solar-scaled models from \cite{allende-prieto+2018}. 

Metallicity and effective temperature errors were increased following the prescription of \citet[see their Section~5.1]{husser+2016}. In addition, we added $1\,\rm km\,s^{-1}$ in quadrature to our velocity uncertainties to account for the uncertainty in the MUSE wavelength calibration (\citealt{kamann+2016,kamann+2018, weilbacher+2020}). Indeed, velocity uncertainties measured by \texttt{SPEXXY} have been proven to be a reliable description of the true errors, although they could be slightly underestimated (overestimated) for stars with lower (higher) $S/N$ (see \citealt{kamann+2016, kamann+2018}). Stars observed multiple times would allow quantification of an additional correction factor to be added to the measurement error budget (see \citealt{kamann+2016}). Unfortunately, such calibration was not possible with our observations.

\begin{figure}
    \centering
    \includegraphics[width=0.8\columnwidth]{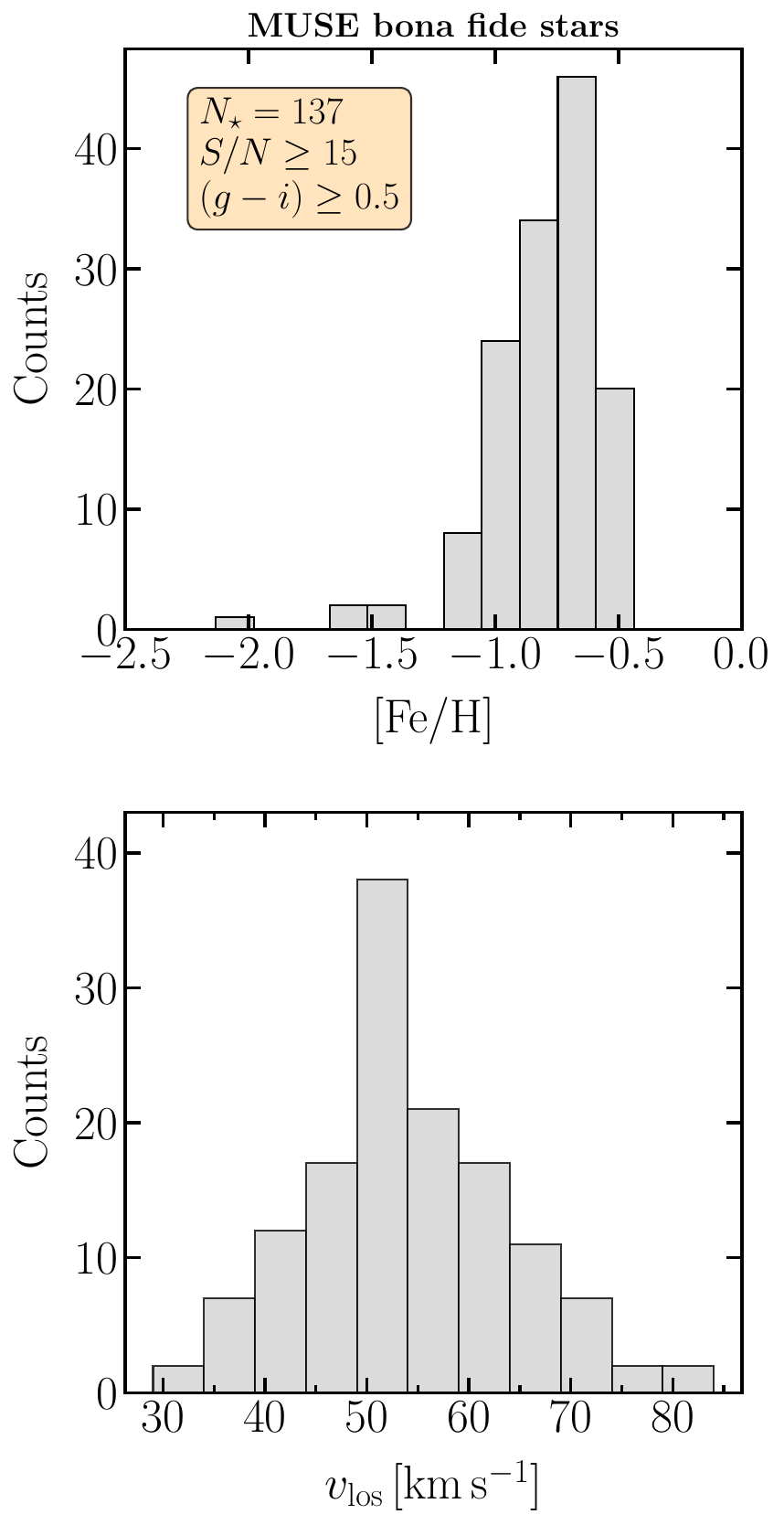}
    \caption{Metallicity (upper panel) and line-of-sight velocity (lower panel) distributions of MUSE bona fide stars created by assuming a bin size of $0.15\,$dex and $5\,\rm km \, s^{-1}$, respectively. The text box displays the final number of stars and summarises the cuts applied.}
\label{fig: spexxy_met_vel_dist}
\end{figure}

To investigate the chemo-kinematical properties of our sample, we selected stars with $S/N\geq 15$ per pixel, which correspond to uncertainties on the velocity and [Fe/H] values measured by \texttt{SPEXXY} smaller than $5\,\kms$ and $0.15\,$dex, respectively. This S/N limit corresponds to an $i$-band magnitude of $\sim22$ (see Fig.\,\ref{fig: pampelmuse_results}).
Moreover, we selected stars with colours $(g-i)\geq0.5$ to exclude any very young MS stars and focus our analysis on RGB and RC objects, whose initial input atmospheric parameters and metallicities are better constrained. After this selection, the final sample within the MUSE FOV contains 137 stars. While this is a relatively small number of spectra compared to other studies \citep[e.g.,][see the discussion of Sect.~\ref{sec: literature_comparison}]{battaglia+2006, pace+2021}, it is the densest sampling of the inner region of the Fornax dSph to date. The metallicity distribution function (MDF) of bona fide stars is shown in the upper panel of Figure~\ref{fig: spexxy_met_vel_dist}, where the bin size is chosen to be $0.15\,$dex. The corresponding $\vlos$ distribution is shown in the bottom panel, plotted by assuming a bin size equal to $5\,\rm km\,s^{-1}$.

\begin{figure*}
    \centering
    \includegraphics[width=0.9\hsize]{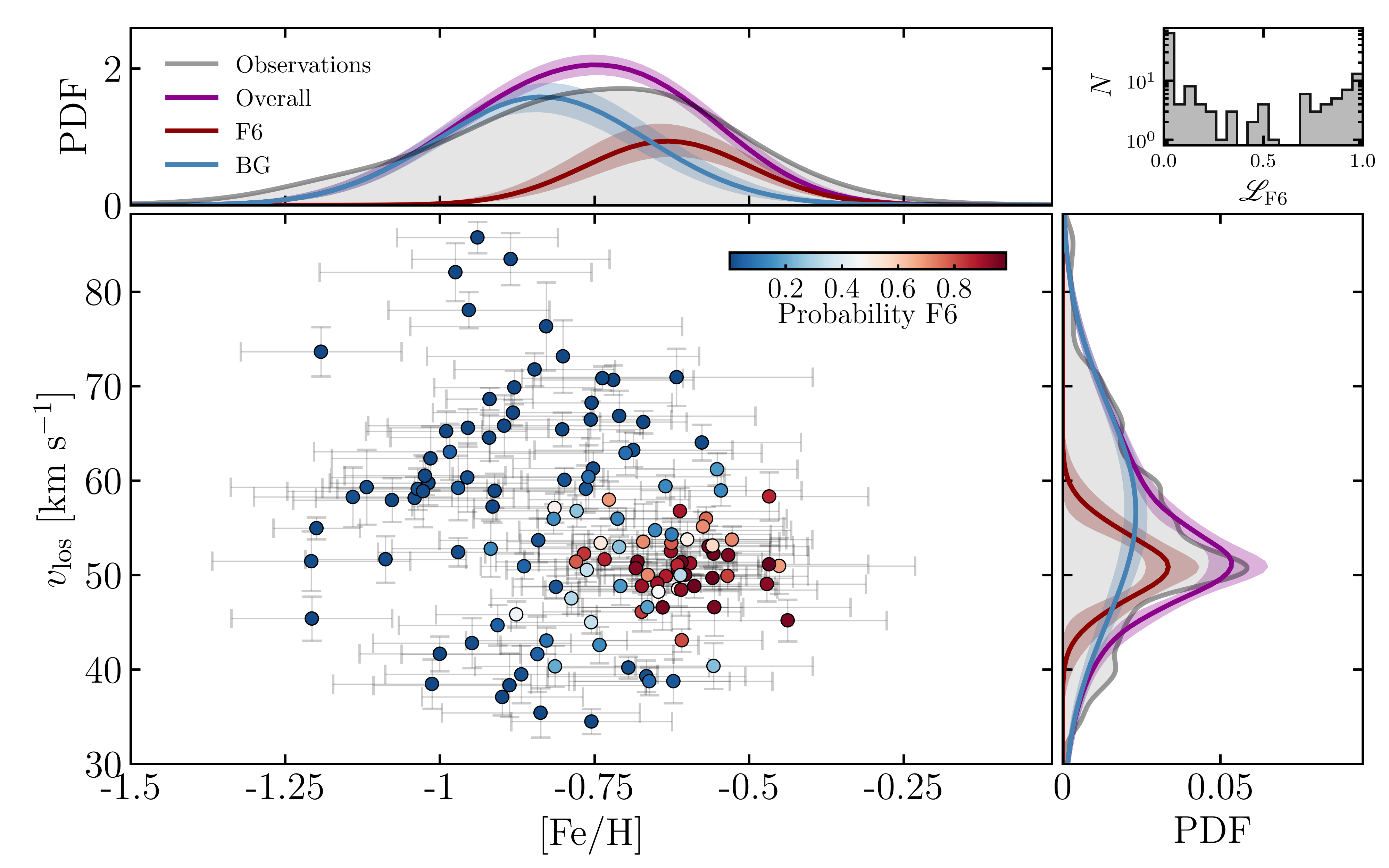}
    \caption{Main panel: 2D distribution of line-of-sight velocities and metallicities for the data MUSE set described in Section~\ref{sec: spectroscopy}. Stars more metal-poor than $\met < - 1.4\,$dex were excluded from the analysis. Each star is colour-coded according to its probability of belonging to Fornax~6, as defined in \ref{subsec:bayes}, while grey error bars represent the uncertainties on the measurements, as computed by \texttt{SPEXXY}. Red and blue colours correspond to high probabilities of belonging to the Fornax~6 cluster or to the background component, respectively. Top and right panels: Median marginalised distributions of metallicity (top) and line-of-sight velocity (right) are shown as solid coloured lines and are compared to the observed distributions (grey solid curve). The contribution of the Fornax~6 overdensity and the background are shown in red and blue, respectively, while their sum is shown in purple. Shaded bands indicate the $1\sigma$ confidence intervals. The observed distributions were built by smoothing the data using their individual measurement uncertainties, with each data point represented by a Gaussian kernel. The top-right panel shows the probability of belonging to the Fornax~6 overdensity for all stars in the sample.}
    \label{fig:met_vel_dist}
\end{figure*}

\section{Characterisation of Fornax~6}
\label{sec:charac}

To extract the morphological and chemo-kinematical properties of Fornax~6, we employed a parametric classification scheme based on a mixture modelling approach, in which stars are probabilistically assigned to different components by maximising the likelihood of the given data set. The input data may comprise any combination of positions on the plane of the sky,  kinematics, and/or metallicities.  The method fits individual stars within a full Bayesian framework, avoiding arbitrary data binning. The approach is fully described in Appendix~\ref{subsec:method}, and a similar methodology has been applied in several studies (see, e.g., \citealt{walker_penarrubia2011, Martin2016, pace+2020, smith+2023, pascale+2026}).

We considered a two-component model in which stars can be assigned to either Fornax~6 (F6) or the dSph field population, referred to as background (BG). As described in the introduction, Fornax has been reported to host three distinct chemo-dynamical populations (see Sect.~\ref{sec: introduction}). However, given the small FOV of MUSE relative to the size of the galaxy, we approximated the background using a single chemo-kinematic component.
Following previous work \citep{wang+2019,pace+2021}, we considered a flattened Plummer model \citep{Plummer1911} to describe the spatial distribution of the Fornax~6 stars, while we assumed that the density of the Fornax field stars is constant over the observed FOV.  
Metallicity and velocity distributions were modelled with Gaussian functions.

We adopted uniform priors on the model's free parameters and sampled from the posterior distribution using a Markov Chain Monte Carlo (MCMC) method (see Appendix~\ref{subsec:method} for details). The fitting procedure returns the central coordinates of Fornax~6 ($ \xc, \yc$), defined as offsets with respect to the centre measured in \cite{wang+2019}, half-light radius\footnote{We stress that the method returns the half-density radius. However, we can compare the two quantities assuming that stars contribute equally to the system's luminosity.} ($\Rh$), ellipticity ($e$), position angle ($\phi$), mean metallicity and metallicity dispersion ($\overline{\met}, \sigma_\met$), systemic velocity and velocity dispersion ($\overline{v_{los}}, \sigma_{v}$), and the fraction of stars of the input catalogue associated to Fornax~6 ($w$).
Median parameters and relative $1\sigma$ confidence intervals were defined as the 50th, 16th and 84th percentiles of the one-dimensional marginalised posterior distributions.

We adopted two complementary approaches based on the different data sets. Firstly, we applied the method to the MUSE data set, which enables a complete chemo-kinematical-spatial characterisation of the cluster (Section~\ref{sec: chemo-kinematical_fit}). 
Secondly, we applied a modified version of the method to the GMOS photometry, focusing only on the spatial characterisation of Fornax~6 (Section~\ref{sec: structural_properties}). In this case, the method is very similar to that presented in \citet[][see also \citealt{smith+2023, bellazzini+2026}]{Martin2008, Martin2016}. Our goal here was to robustly assess the presence and properties of the Fornax~6 overdensity by exploiting a much deeper photometric data set. Building upon the results in Section~\ref{sec: structural_properties}, we estimated the total luminosity of the cluster. 

The measurements derived from the two approaches are summarised in Table~\ref{tab:newparams2}, together with the adopted priors. A summary of the newly determined properties of Fornax~6 is presented in Table~\ref{tab: F6}.

\subsection{Chemo-kinematical properties}
\label{sec: chemo-kinematical_fit}

\begin{figure*}
    \centering
    \includegraphics[width=0.9\hsize]{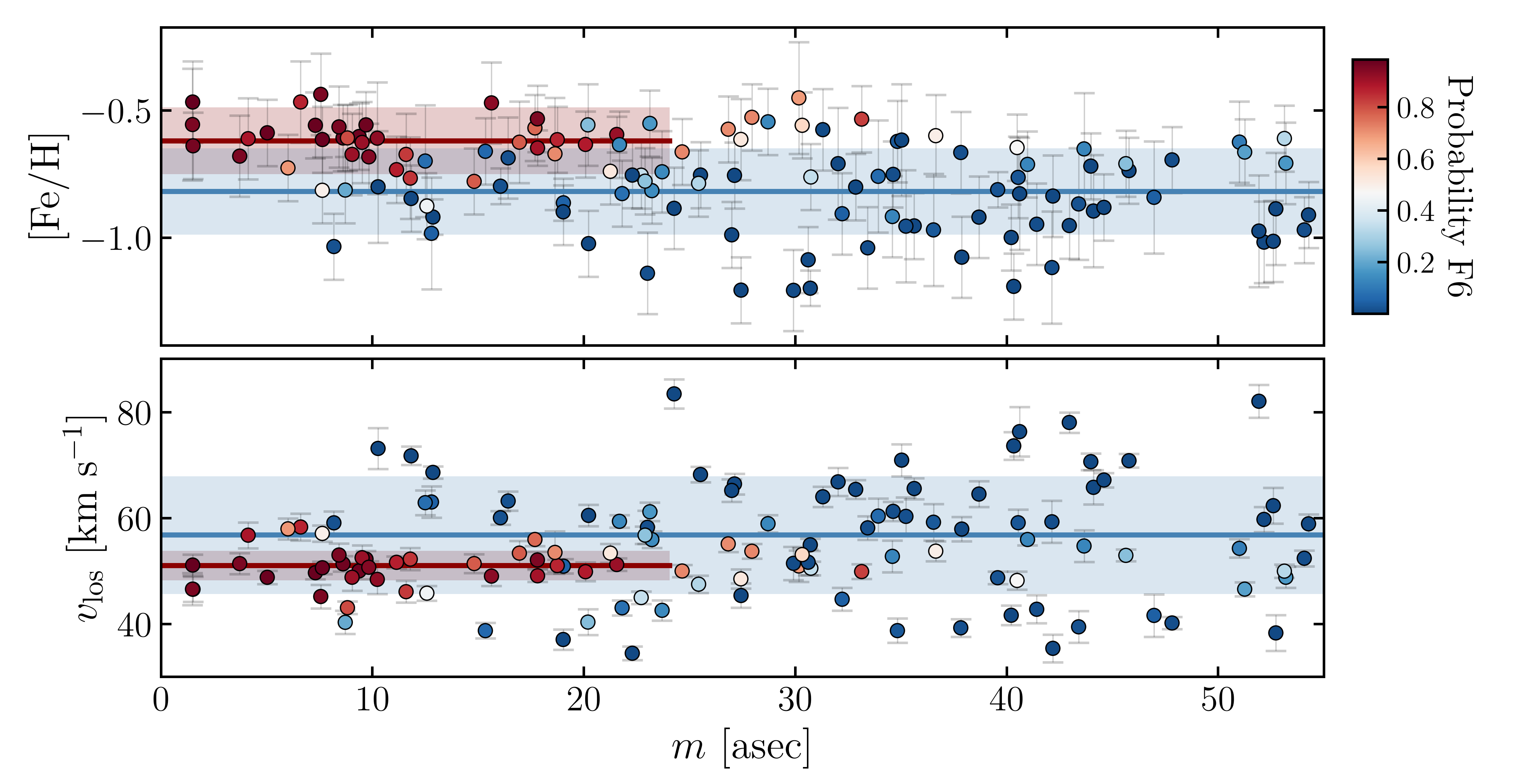}
    \caption{Top panel: Position-metallicity diagram for the data set described in Section~\ref{sec: spectroscopy}. Each star is colour-coded according to its probability of membership in the Fornax~6 cluster. Grey error bars represent the uncertainties on the measurements as computed by \texttt{SPEXXY}. Red and blue colours indicate high probabilities of belonging to the Fornax~6 cluster and the background component, respectively. Bottom panel: Same as the top panel, but showing the position-velocity diagram. The solid coloured lines correspond to the mean metallicity and line-of-sight velocity distributions from the models with the highest posterior probability among those explored by the MCMC. The bands represent a $\pm1\sigma$ range, where $\sigma$ is computed by adding in quadrature the median observational error and the dispersion of the corresponding distribution taken from the model with the highest posterior.}
    \label{fig:structchemodyn}
\end{figure*}

The fit was performed on 131 stars from the bona fide spectroscopic catalogue, after excluding five outlying stars with metallicities $\met < -1.4\,$dex (see Fig.~\ref{fig: spexxy_met_vel_dist}) and one star not satisfying the CMD colour selection mentioned in Section~\ref{sec: photometric_catalogue}. The model is described by equation~\ref{for:pfnx}, with the likelihood function defined as in equation~\ref{for:like}.
The marginalised two- and one-dimensional posterior distributions, reported in Appendix~\ref{app:appA}, provide a detailed view of the parameter correlations and show that all model parameters are well constrained. Median values and associated uncertainties are reported in Table~\ref{tab:newparams2}.

%   met vel plane
Figure~\ref{fig:met_vel_dist} shows the distribution of stars in the [Fe/H] versus line-of-sight velocity plane. Each point is colour-coded according to the probability of membership of the Fornax~6 overdensity (as derived from Eq.~\ref{for:member}), with red and blue indicating high and low probability, respectively. This reflects the membership probability distribution shown in the small inset in the upper-right corner of the plot.
The high-probability Fornax~6 members clearly cluster in the [Fe/H]-$\vlos$ space, with a mean metallicity of $-0.63_{-0.03}^{+0.03}\,$dex and a mean line-of-sight velocity of $50.8_{-0.7}^{+0.8}\,$km s$^{-1}$ (see also Table~\ref{tab:newparams2} and Figure~\ref{fig:cornerplot_MUSE}). The number of stars statistically associated with the cluster is $43\pm 10$. 

The top and right panels of Figure~\ref{fig:met_vel_dist} show the marginalised one-dimensional distributions of metallicity and line-of-sight velocity, respectively. The solid coloured curves represent the median model for each component, while the solid purple curve represents the median total model. The models have been convolved with the median observational uncertainties in metallicity and velocity (0.13 dex and 1.9 km s$^{-1}$, respectively) to enable a direct comparison with the data. 
We note that the inferred metallicity dispersion of Fornax~6 ($\sigma_{\met,\fnx}<0.04\,$dex, 68\% upper limit) is smaller than the width of the corresponding Gaussian component shown in the top panel of Figure~\ref{fig:met_vel_dist}, hence is unresolved. This is because the observed width of the MDF is dominated by measurement uncertainties, which are much larger than the cluster's intrinsic dispersion.

Figure~\ref{fig:structchemodyn} presents the distribution of stellar metallicity (top panel) and line-of-sight velocity (bottom panel) as a function of elliptical radius $m$ (Eq.~\ref{for:m}). For each star in the sample, $m$ is computed using the set of ellipse parameters corresponding to the model with the highest posterior from the MCMC fitting (see red squares in Fig.~\ref{fig:cornerplot_MUSE}). As for Figure~\ref{fig:met_vel_dist}, stars are colour-coded according to their membership probability.
The two populations occupy distinct regions in the position–metallicity and position–velocity diagrams, with a notable difference also in their spatial extent: stars associated with the candidate cluster are visibly more concentrated ($m\lesssim10\arcsec$), and exhibit smaller dispersions in both metallicity and line-of-sight velocity than stars belonging to the background component. This is particularly evident in the metallicity panel, where cluster members form a well-defined horizontal sequence around $\overline{\met}_{\fnx}$, clearly offset from the background population. To highlight these trends, we overlay shaded horizontal bands centred on the mean metallicity and velocity of the two components, whose widths were computed by adding the median observational error and the intrinsic dispersion of the corresponding distribution in quadrature. These mean values and dispersions are taken from the highest posterior model.

\begin{figure*}
    \centering
    \includegraphics[width=1\hsize]{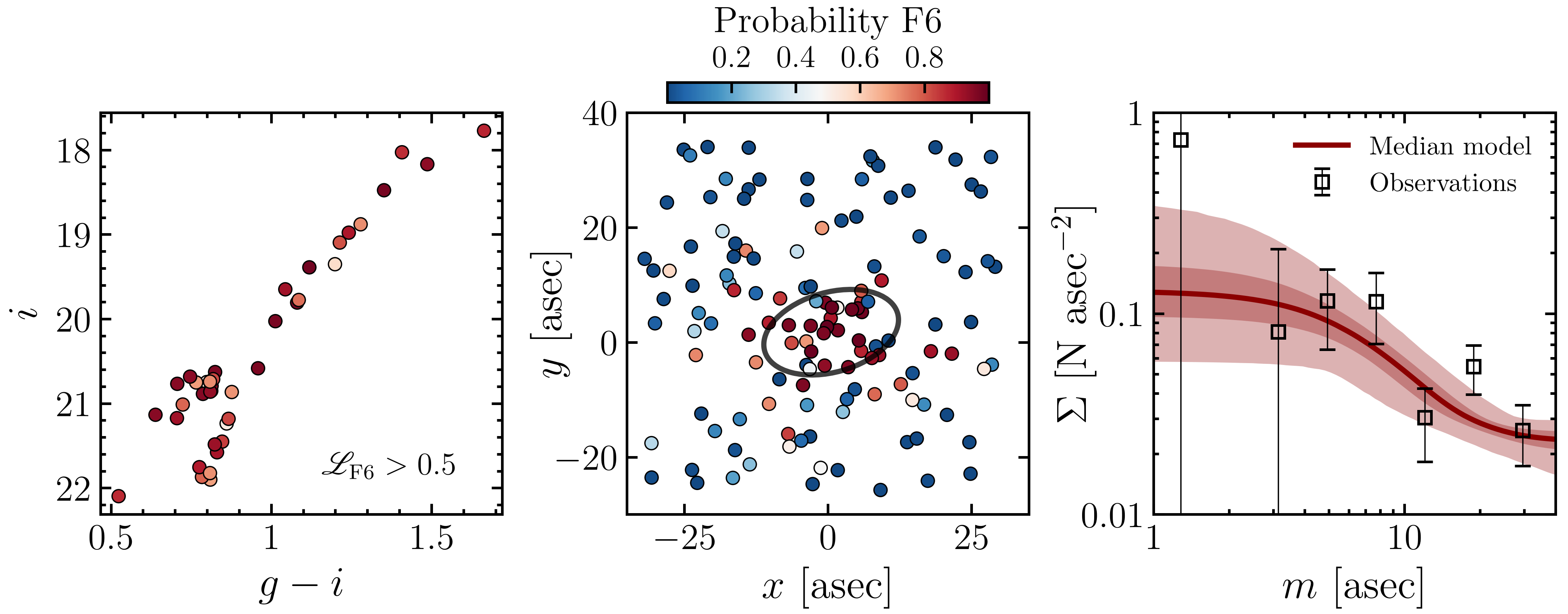}
    \caption{Left panel: CMD showing GMOS $(g - i)$ versus $i$ photometry for stars in the MUSE catalogue, colour-coded by their probability of belonging to the Fornax~6 cluster. For clarity, only stars with a membership probability $\Lprob_{\fnx} > 0.5$ are shown; stars below this threshold, likely members of the background population, are omitted. Central panel: Spatial distribution of stars in the MUSE catalogue, colour-coded by their membership probability in Fornax~6. The solid black ellipse is centred on the Fornax~6 centroid and has a semi-major axis equal to the system’s half-mass radius. The parameters used to draw the ellipse correspond to the model with the highest posterior probability among those explored by the MCMC. Right panel: Median model surface density profile along the system’s semi-major axis (solid red line), with $1\sigma$ and $3\sigma$ confidence intervals (red bands), compared to the binned surface density profile of the cluster (black squares with error bars).}
    \label{fig:spatial_structchemodyn}
\end{figure*}

%   cmd & structural properties
Finally, Figure~\ref{fig:spatial_structchemodyn} provides an overview of the photometric and structural properties of the Fornax~6 cluster inferred from the fit to the MUSE catalogue. The left panel shows the CMD in the ($g-i$) versus $i$ bands. For the sake of clarity, even though stars are colour-coded for the probability to belong to the Fornax~6 cluster, we only show those with probability membership $\Lprob_{\fnx}>0.5$. These stars follow a well-defined cluster-like sequence, with no dependence of cluster membership probability on stellar luminosity. 
The central panel of Figure~\ref{fig:spatial_structchemodyn} shows the two-dimensional spatial distribution on the plane of the sky. A prominent overdensity of likely members is evident in the central region. A solid black ellipse is overplotted, with a semi-major axis corresponding to the elliptical half-light radius derived from the model with the highest posterior explored by our MCMC analysis. %Additionally, we overlay a dashed ellipse deriving from the analysis of the stand-alone GMOS catalogue, which is $\sim$3 magnitudes deeper (see Section~\ref{sec: structural_properties}). The two ellipses are in excellent agreement, reinforcing the robustness of our measurements across independent data sets. 
The right panel of Figure~\ref{fig:spatial_structchemodyn} shows the median radial density profile of Fornax~6 from the fit (red line), and the density profile computed from the data (black squares). 

The observed binned density distribution (see the black squares in Fig.~\ref{fig:spatial_structchemodyn}) was constructed using logarithmically spaced elliptical bins, whose ellipticity and position angle were determined through a non-parametric procedure. This approach involves the diagonalisation of the shape tensor of the cluster on the plane of the sky (\citealt{Zemp2011}; see also \citealt{dalessandro+2024} for a full description of the method and application to a stellar cluster).
In this way, we can compare the model surface density distribution with the distribution directly inferred from the data.
Due to low number statistics, stars were divided into three radial macro-bins, each with a width equal to $\Rh$. We then iteratively diagonalised the shape tensor of the stellar positions in each macro-bin until convergence to the 1\% level was achieved (see, e.g., \citealt{dalessandro+2024, dellacroce+2025}). The flattening and position angle assumed for each bin illustrated in Figure~\ref{fig:spatial_structchemodyn} (right panel) depend on the macro-bin they fall into.
Uncertainties on the observed density profile were defined as the dispersion among the profiles computed by dividing the space into three angular subsectors.

\begin{table}
\begin{minipage}{0.48\textwidth}
  \centering
    \caption{Summary of parameter values resulting from the fit of the MUSE and the GMOS catalogues in Section~\ref{sec: chemo-kinematical_fit} and \ref{sec: structural_properties}, with the method described in Appendix~\ref{subsec:method}. The columns, from left to right, list the parameter name, the assumed prior, and the median parameter values with $1\sigma$ confidence intervals. The term $Nw$, in which $N$ represents the total number of sources in the input catalogue, estimates the number of stars belonging to the Fornax~6 cluster.}
    \renewcommand{\arraystretch}{1.75}
    \begin{tabular}{lcll}
        \toprule
        \textbf{Parameter} & \textbf{Prior} & \textbf{F6\_MUSE}  & \textbf{F6\_GMOS}\\
        \bottomrule
        $\xc$ (asec) & [-10, 10] & $-0.04^{+2.06}_{-2.18}$ &   $1.79^{+0.85}_{-0.82}$  \\
        $\yc$ (asec) & [-10, 10] & $1.55^{+1.52}_{-1.68}$ &    $3.84^{+0.98}_{-1.00}$  \\
        $\Rh$ (asec) & [0, 60] & $13.8^{+4.8}_{-3.6}$  &    $12.37^{+1.48}_{-1.79}$  \\
        $\Rh$ (pc) & - & $9.6^{+3.3}_{-2.5}$  &    $8.6^{+1.0}_{-1.2}$  \\
        $e\equiv1-q$   & [0,1] & $<0.47$\footnote{\label{note_tab1}68\% upper limit.}              &    $<0.26$\footref{note_tab1} \\
        $\phi$ (deg)      & [0,180] & $99.9^{+23.4}_{-37.1}$         &    $58.7^{+103.9}_{-44.0}$ \\
        $\overline{v_{los}}_{,\fnx}$ $(\kms)$ & [30, 90] &  $50.8_{-0.7}^{+0.8}$    &  -   \\
        $\sigma_{v,\fnx}$ $(\kms)$ & [0, 30] &  $3.3_{-1.9}^{+1.4}$ &  -  \\
        $\overline{\met}_\fnx$ (dex) & [-0.8, -0.3] & $-0.63_{-0.03}^{+0.03}$ &     -    \\
        $\sigma_{\met,\fnx}$ (dex) & [0, 1] & $<0.04$\footref{note_tab1} &   -     \\
        $\overline{v_{los}}_{,\BG}$ $(\kms)$ & [30, 90] & $56.6_{-1.3}^{+1.4}$  &      -       \\
        $\sigma_{v,\BG}$ $(\kms)$ & [0, 30] & $11.4_{-0.9}^{+1.0}$ &   -      \\
        $\overline{\met}_\BG$ (dex) & [-1.2, -0.7] &  $-0.83_{-0.03}^{+0.02}$ &   -       \\
        $\sigma_{\met,\BG}$ (dex) & [0, 1] & $0.11_{-0.02}^{+0.02}$ &   -      \\
        $w$ & [0, 1] & $0.32^{+0.08}_{-0.08}$ &      $0.043^{+0.006}_{-0.005}$     \\
        $Nw$ & - & $43_{-10}^{+10}$ &            $370^{+51}_{-43}$          \\

        \bottomrule
    \end{tabular}
    \label{tab:newparams2}
\end{minipage}
\end{table}

\subsection{Structural properties }  
\label{sec: structural_properties}

%   intro 

%   about the data set 
The fit was made to the culled GMOS photometric catalogue, described in Section~\ref{sec: photometric_catalogue} (see the black CMD of Fig.~\ref{fig: cmd_density}). To avoid effects due to detector defects and density gradients, which are visible in the density maps in the magnitude range $23<i<27$ (Fig.~\ref{fig: cmd_density}, right panel), we selected stars within a circular region of radius $R=73\arcsec$ centred on the cluster's literature coordinates \citep[see their Table~1]{wang+2019}. We further restricted to stars brighter than $i= 25.5$, as this cut is one magnitude brighter than the observed turnover of the $i$-band luminosity function, minimising potential issues due to varying incompleteness. 
%However, as a further check, we have verified that the fit yields consistent results within $1\sigma$ across different magnitude cuts (see Appendix?). Finally, we assumed a negligible error in the stars' positions.
Median values and errors of the model parameters are listed in the rightmost column of Table \ref{tab:newparams2}, while the resulting marginalised one- and two-dimensional posterior distributions can be found in Appendix~\ref{app:appA} (Fig.~\ref{fig:cornerplot_GMOS}).  We note that the measurements from the two data sets are consistent within $1.5\sigma$, and uncertainties from fitting the GMOS catalogue are smaller, except for $\phi$.

\begin{figure*}
    \centering
    \includegraphics[width=0.8\linewidth]{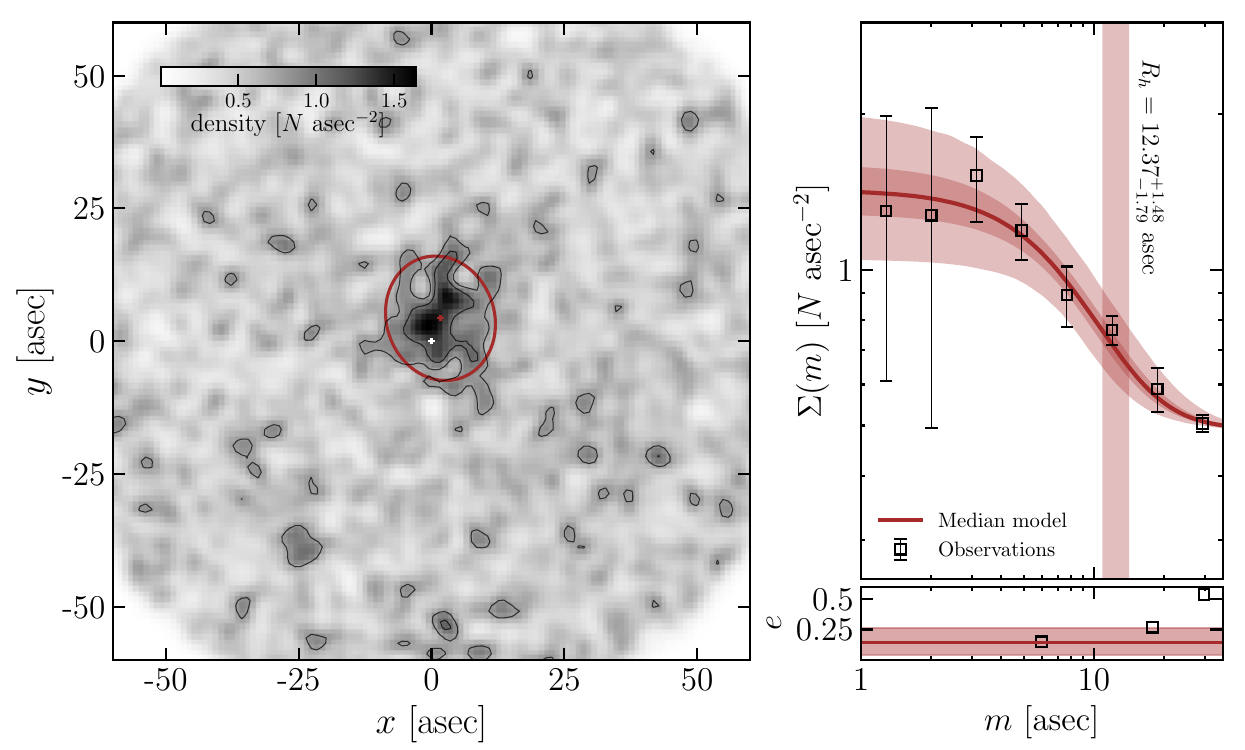}
    \caption{Left panel: projected stellar density distribution, based on GMOS, in the region of the sky used for the fit in Section~\ref{sec: structural_properties}. To enhance the overdensity associated with the Fornax 6 cluster, stars have been binned into 1 asec pixels, and the resulting map has been smoothed using a Gaussian kernel with a 1.5 asec dispersion. The newly determined Fornax~6 centre, as derived by our fit, is marked with a red cross, while the previous literature estimate (from \citealt{wang+2019}) is indicated by the white cross.
    The red elliptical curve represents the system's effective radius. The parameters used to construct the model in the left panel correspond to those of the highest-posterior solution among the explored models (see Appendix~\ref{subsec:method} for details). The black isodensity contours correspond to densities of 6 and 8 times the background noise level, computed as the standard deviation in the region $25 \arcsec \le x \le 50 \arcsec$ and $25 \arcsec \le y \le 50 \arcsec$. Top-right panel: median surface density distribution along the system's semi-major axis (solid red line), with $1\sigma$ and $3\sigma$ confidence intervals (red bands), compared to the binned density distribution (black squares with error bars). The red vertical band indicates the $1\sigma$ confidence range for the system's effective radius. Bottom-right panel: ellipticity from the median model (red solid line) with the $1\sigma$ confidence interval compared to a non-parametric estimate of the cluster ellipticity obtained by diagonalising the image shape tensor (black squares).}
    \label{fig:spatialdistribution}
\end{figure*}

Figure~\ref{fig:spatialdistribution} provides an overview of the structural properties of the cluster based on our fit, and it compares them with observations. The left panel displays the 2D stellar density distribution in the relevant region of the sky used for the fit. To highlight the cluster, we drew two black isodensity contours corresponding to densities 6 and 8 times the background noise level, estimated in a square region far from the cluster centre ($25 \arcsec \le x \le 50 \arcsec$ and $25 \arcsec \le y \le 50 \arcsec$). The newly measured centre is marked with a red cross, while the red elliptical contour measures the half-mass radius of the system. The centre, effective radius, ellipticity, and position angle used to draw the red elliptical contour correspond to those of the model with the highest posterior among those explored by the MCMC (see red squares in Fig.~\ref{fig:cornerplot_GMOS}). %The position of this model is indicated by red squares in the posterior distribution of Figure~\ref{fig:posterior_fit}. 
The elongation and size of the model provide a close match to the underlying stellar overdensity in Figure~\ref{fig:spatialdistribution}. However, the emergent morphology is clearly irregular.

% comment to right panels 
In the right panels of Figure~\ref{fig:spatialdistribution}, we compare the surface density distribution of the model with the one inferred from the data. The top-right panel displays the surface density profile along the system’s semi-major axis, as obtained from the fit, compared to the binned density distribution of the cluster (black squares with error bars) calculated using the same non-parametric method described in the previous section (Sect.~\ref{sec: chemo-kinematical_fit}).
In addition, the three ellipticity values derived by diagonalising the shape tensor are shown in the bottom-right panel. Within $2\Rh$, these values exhibit excellent agreement with the results from the Plummer fit, further confirming that the parametric model provides a good representation of the data. 
We note that both $\Rh$ and $e$ determined in this work are significantly smaller than the measurements made by \cite{wang+2019}, who made use of a shallower photometric catalogue derived from lower spatial resolution images.

To estimate the total luminosity of Fornax~6 without making an assumption about the properties of the stellar population, we performed aperture photometry on the stacked GMOS images using the python library \texttt{photutils} \citep{photutils}. Firstly, we subtracted a low-resolution 2D background image from the data, generated by the task \texttt{Background2D}. Specifically, we divided the input image into cells of $30\times30\,\rm pixel^2$, and derived the background map by interpolating the $3\sigma$-clipped background value of each box cell obtained with the \texttt{MedianBackground} class, assuming \texttt{filter\_size} $=3$.
Secondly, we estimated the contribution of the stellar background by calculating the mean value of a circular annulus of inner radius $3\Rh$ and outer radius $5\Rh$ from the centre of the overdensity, after having performed an iterative $3\sigma$-clipping procedure.
We also masked the four bright background galaxies visible along the line of sight to Fornax~6.  The total magnitude of the cluster was assumed to be the background-subtracted pixel sum within a radius of $2\Rh$.

Finally, we derived the absolute $M_g$ and $M_i$ magnitudes assuming the distance and associated error to the Fornax dSph from  \cite{oakes+2022}, and a reddening value of $E(B-V) = 0.06$ (see Section~\ref{sec: spectroscopy}). We obtained the absolute magnitude in the $V$-band using the transformation equation provided for the DES DR2 survey\footnote{https://des.ncsa.illinois.edu/releases/dr2/dr2-docs/dr2-transformations}\citep{abbott+2021}. The resulting absolute magnitude and total uncertainty are $M_V = -5.0 \pm 0.4$. Adopting the absolute AB $V$-band magnitude of the Sun \citep{willmer2018}, this translates into a total $V$-band luminosity of $L_V = 8.3 \pm 3.1 \times 10^3\, L_{\sun}$. This is in agreement with \cite{wang+2019}, who estimated $M_V = -4.8 \pm 0.4$ by converting the number of stars associated with Fornax~6 into luminosity, assuming an isochrone model with $t =10\,$Gyr and $\met=-1.5\,$dex.

\begin{table}
\begin{minipage}{0.48\textwidth}
  \centering
    \renewcommand{\arraystretch}{1.75}
\caption{Overall summary of the properties of Fornax~6 derived in this paper. Structural properties and absolute magnitude are based on GMOS data, while chemo-kinematical properties were derived from MUSE.}
    \begin{tabular}{lr}
        \hline
        \textbf{Parameter} & \textbf{Value}  \\
        \hline
        $\alpha$ (deg)  &  $40.0281\pm 0.0002$  \\
        $\delta$ (deg) &   $-34.4209\pm 0.0003$  \\
        $\Rh$ (asec)  &  $12.37^{+1.48}_{-1.79}$ \\
        $\Rh$ (pc) &   $8.6^{+1.0}_{-1.2}$ \\
        $e$    &  $<0.26$\footnote{\label{note_tab2}68\% upper limit.} \\
        $\phi$ (deg) & $58.7^{+103.9}_{-44.0}$  \\
        $M_V$ (mag) &  $-5.0\pm 0.4$ \\
        $\vlos$ $(\kms)$ & $50.8^{+0.8}_{-0.7}$ \\
        $\sigma_v$ $(\kms)$ & $3.3^{+1.4}_{-1.9}$  \\
        $\met$ (dex) & $-0.63\pm0.03$ \\
        $\sigma_{\met}$ (dex) & $<0.04$\footref{note_tab2} \\
        \bottomrule
\end{tabular}
\label{tab: F6}
\end{minipage}
\end{table}

\section{Discussion}
\label{sec: discussion}

\subsection{Comparison with literature studies}
\subsubsection{Chemo-kinematics in the central region of Fornax}
\label{sec: literature_comparison}

Several spectroscopic studies have mapped the complex chemo-kinematical properties of the Fornax galaxy over the last 20 years. However, only a few have sampled the inner region of the dSph, where Fornax~6 resides. As a reminder, our MUSE field is centred at a projected radius of $0.27\,$kpc and covers a FOV of 1 square arcmin.

\cite{battaglia+2006} presented CaII triplet metallicities and line-of-sight velocities obtained with VLT-FLAMES for more than 550 member RGB stars. We cross-matched our MUSE catalogue with their catalogue, retrieved from the Strasbourg Astronomical Data Centre (CDS), and found only one star in common. This is not unexpected given their observations came from the wide-field GIRAFFE/MEDUSA fibre spectrograph and did not focus specifically on Fornax's central region; indeed, only 73 members from their sample lie within the central $350\,$pc. Nonetheless, we found good agreement for this star, with metallicity agreeing within $0.1\,$dex.
\cite{battaglia+2006} split their catalogue into two metallicity components, identifying metal-rich stars as those with $\rm [Fe/H]>-1.3\,dex$, and they studied the velocities of this component within three radial bins. In the innermost bin ($r<0.4\degree = 1\,$kpc), they measured  $\overline{v_{los}} = 53.3\pm0.8\,\rm km\,s^{-1}$ and $\sigma_{v} = 11.3 \pm 0.5\, \rm km\,s^{-1}$. Comparing these values to those of our background component, we find good agreement with the velocity dispersion ($\sigma_{v,\BG} = 11.4_{-0.9}^{+1.0}\,\kms$), while we find a slightly discrepant mean velocity ($\overline{v_{los}}_{,\BG}=56.6_{-1.3}^{+1.4}\,\kms$).

Later, \cite{amorisco&evans2012} used the data sets of  \cite{walker+2009} and \cite{battaglia+2006} to argue for three chemo-dynamical components. Using a Bayesian approach to jointly fit the chemical and kinematical information, they found statistical evidence for dividing the metal-rich stars into two subpopulations with mean metallicities of $\rm [Fe/H] = -0.95\,dex$ and $\rm [Fe/H] = -0.65\,dex$. The more metal-rich stars were found to be more centrally concentrated, spanning a radial range that also includes Fornax~6, and characterised by colder kinematics. Specifically, \cite{amorisco&evans2012} measured $\sigma_{v} = 11.3 \pm 0.7\, \rm km\,s^{-1}$ and $\sigma_{v} = 8.6\pm1\, \rm km\,s^{-1}$ for the intermediate and metal-rich component, respectively (mean line-of-sight velocities were not provided). The metallicities derived for these two chemo-kinematical subpopulations are very close to those that we measured for stars associated with the background ($\overline{\met}_\BG = -0.83\,$dex) and the cluster ($\overline{\met}_\fnx = -0.63\,$dex). On the other hand, while the $\sigma_v$ of the metal-intermediate population agrees with the one measured for our BG component, the $\sigma_v$ of the metal-rich population differs by almost $3\sigma$ from the velocity dispersion retrieved for Fornax~6 by our analysis ($\sigma_{v,\fnx} = 3.3_{-1.9}^{+1.4}\,\kms$). This supports the hypothesis that Fornax~6 is a distinct cluster of stars associated with the galaxy's innermost stellar population.

Our chemo-kinematic estimates are also in good agreement with the study of \citealt{letarte+2010}, who analysed 81 high-resolution FLAMES spectra of RGB stars in the inner $1\,$kpc of Fornax. However, there are no stars in common with their catalogue. Nonetheless, their metallicity distribution was found to peak at $\met = -0.8\,$dex, although being skewed towards more metal-rich stars. 
The mean line-of-sight velocity measured by \cite{letarte+2010} is equal to $55.9\,\kms$, in good agreement with our BG measurement, while the velocity dispersion is somewhat larger, amounting to $\sigma_{v} = 14.2\,\kms$.

To investigate the nature of Fornax~6, \cite{pace+2021} analysed a large sample of spectroscopic measurements of RGB stars in the direction of the Fornax dSph, obtained via the Magellan M2FS and MMFS multi-object spectrographs. They retrieved almost 3000 stars belonging to the galaxy, but only 23 sources were in the vicinity of Fornax~6. We recover 15 stars in common with them. 
Considering this sample, we measure a small median offset in the metallicity ($\Delta \rm [Fe/H] = -0.21 \pm 0.08\,dex$) and in the line-of-sight velocity measurements ($\Delta v_{\rm los} = 1.0\pm0.8\,\rm km\,s^{-1}$). However, a couple of stars show a substantial difference in metallicity or velocity. 
For stars flagged as Fornax~6 members, \cite{pace+2021} found  $\overline{v_{\rm los}} = 50.5\pm1.7\,\rm km\,s^{-1}$ and  $\sigma_{v} = \SI{5.6(2.0:1.6)}{\rm km\,s^{-1}}$, associated to a metallicity of $\met = -0.71 \pm 0.05$. These values agree within $1\sigma$ with our estimates listed in Table~\ref{tab:newparams2}. We note, however, that only 3 of the stars that \cite{pace+2021} associate with Fornax~6 fall within the $\Rh$ measured in Section~\ref{sec: structural_properties}.

\subsubsection{Background galaxies}
\label{sec: background_galaxies}

\begin{figure}
    \centering
    \includegraphics[width=0.9\columnwidth]{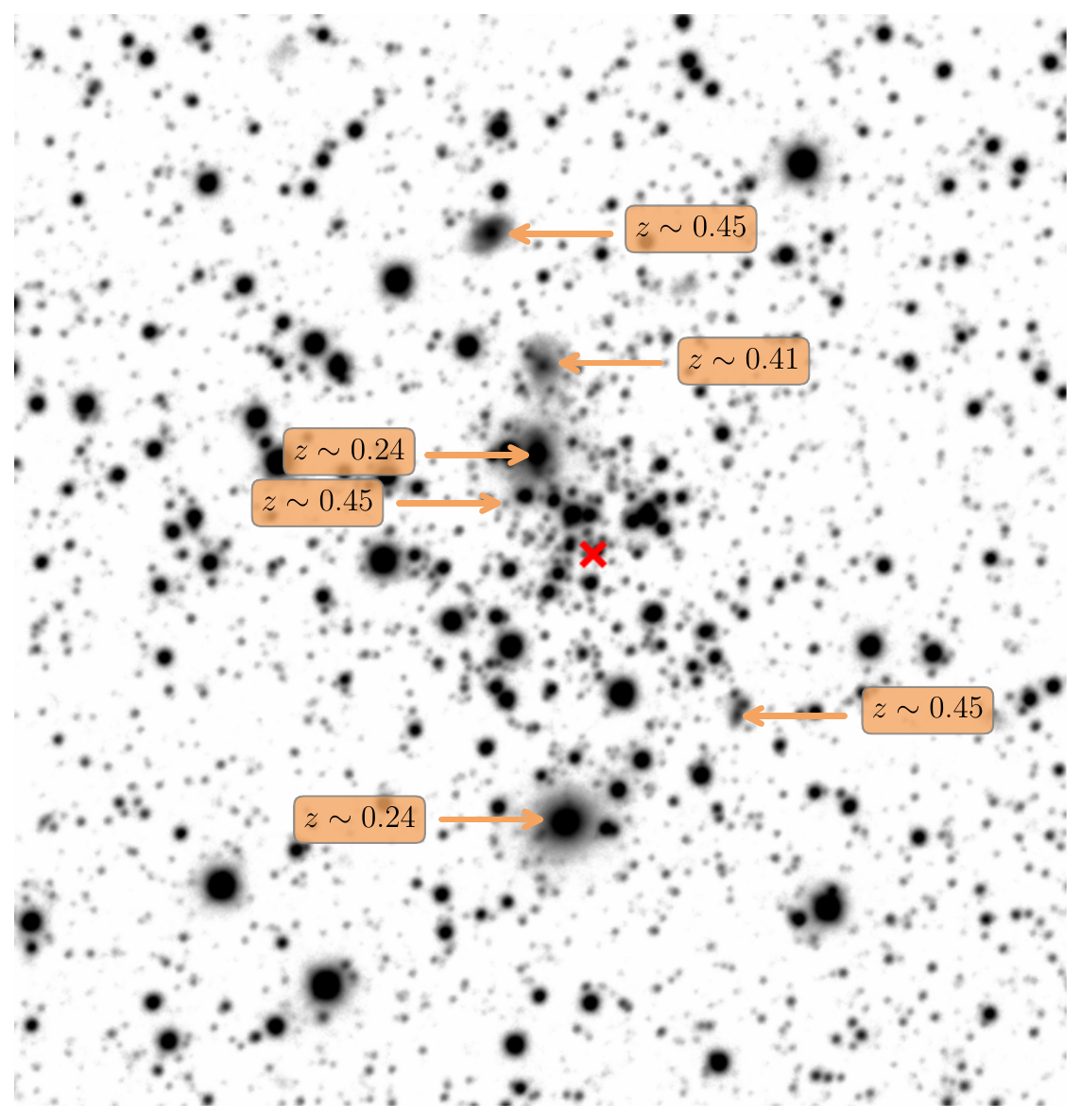}
    \caption{Stacked GMOS $i$-band image showing the location and redshift of some of the background galaxies detected in the MUSE datacube. The centre of Fornax~6, as defined in this study, is marked with a red cross. North is up, east is to the left.}
\label{fig: background_galaxies}
\end{figure}

One of the main historical arguments against Fornax~6 being a star cluster is the
presence of a coincident clustering of background galaxies on the sky \citep[e.g.,][]{stetson+1998}. This can be readily seen in Figure~\ref{fig: background_galaxies}, which shows the stacked $i$-band image from GMOS,  suitably rescaled to highlight the diffuse emission surrounding the background systems.  While the data 
presented in this paper demonstrate a distinct spatial and chemo-kinematical overdensity 
at the position of Fornax~6, they also allow us to explore this population of background galaxies in more detail.

The background galaxies already mentioned in Section~\ref{sec: spectroscopy}, which were found by looking at the extracted sample of MUSE spectra (see star markers in Figure~\ref{fig: pampelmuse_results}), are indicated by the orange arrows in Figure~\ref{fig: background_galaxies}. We also highlighted two additional star-forming galaxies at redshifts comparable to the higher redshift source that were found by inspecting the PSF-subtracted MUSE datacube generated by \texttt{PampelMuse}. Redshifts for the galaxies were derived by calculating the wavelength difference between observed and rest-frame CaII H\&K absorption and [OII] emission lines.

Figure~\ref{fig: background_galaxies} suggests the presence of a grouping of galaxies behind Fornax~6, at $z\sim0.45$. Indeed, by assuming the derived redshifts, we estimate an approximate projected distance between their members of less than $100\,$kpc, suggesting a small compact group. In addition, we also find a similar redshift ($z\sim0.24$) for the two large elliptical galaxies visible in the image.

\subsection{The CMD of Fornax~6 and comparison with the field}  
\label{sec: field_comparison}

\begin{figure}
    \centering
    \includegraphics[width=1\columnwidth]{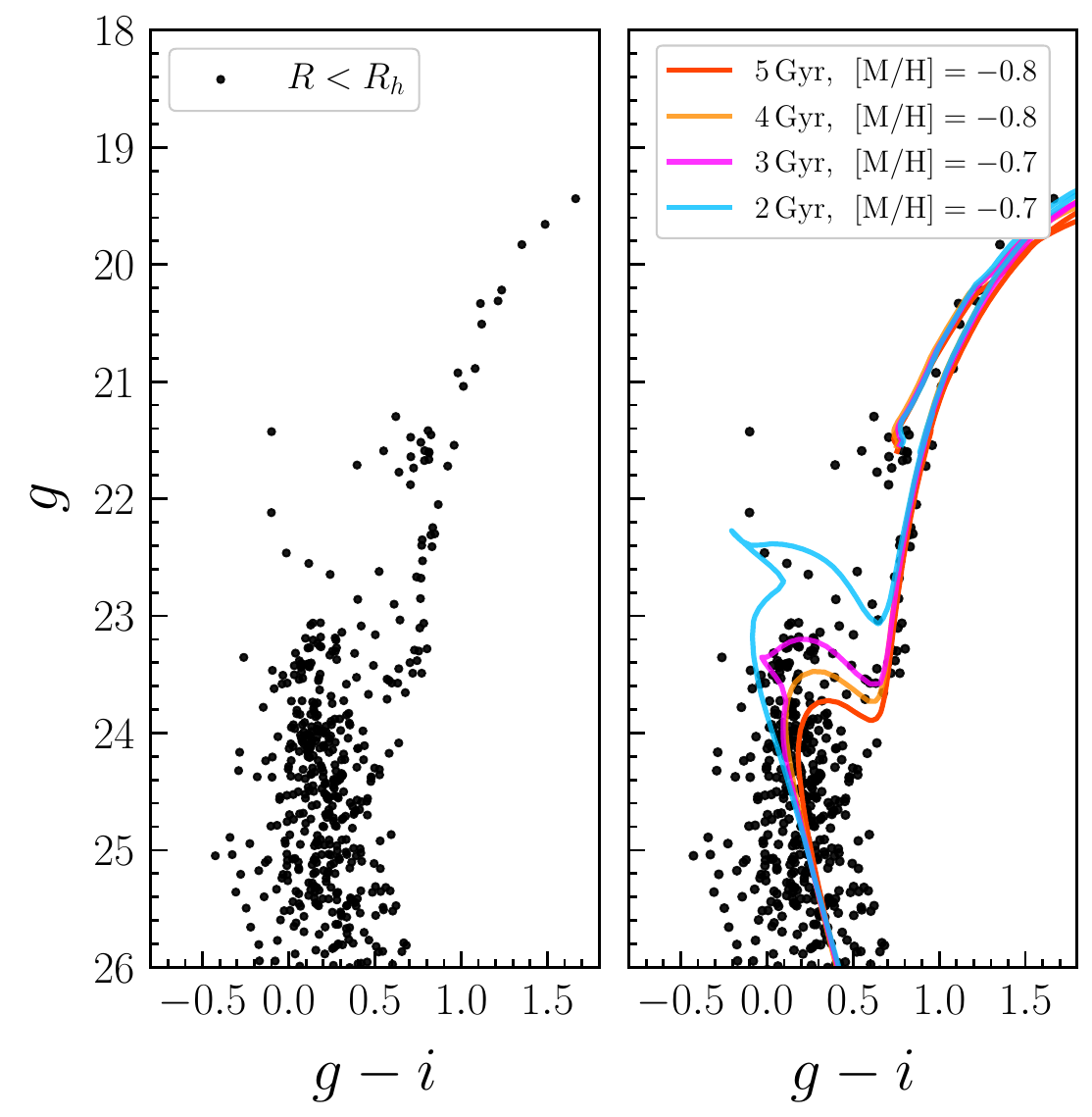}
    \caption{GMOS $(g, g-i)$ CMD of stars within the $R_h$ calculated in this work and centred on Fornax~6 (Section~\ref{sec: structural_properties}). In the right panel, we display, in addition, different PARSEC isochrones calculated using the parameters shown in the legend. Isochrones were reported in the observational plane assuming $E(B-V) = 0.06$ and $\mu = 20.77$ \citep{oakes+2022}.}
\label{fig: CMD_1Rh}
\end{figure}

Having established the reality of the Fornax~6 overdensity, we now focus on its CMD and compare its properties with those of the Fornax dSph field. Our deep GMOS photometry enables us to examine, for the first time, the key region of the MSTO and subgiant branch (SGB), where, in contrast to the RGB, there is greater sensitivity to age and metallicity.  
If Fornax~6 is indeed a stellar cluster, we would expect to observe a well-defined stellar sequence corresponding to a single-age, single-metallicity population.

 We considered stars inside $\Rh$, corresponding to 12.37\arcsec or 8.6\,pc as representative of Fornax~6. We show the respective CMD in Figure~\ref{fig: CMD_1Rh}. In the left panel, we show the $(g, g-i)$ CMD, and in the right panel, we overlay four PARSEC isochrones reported in the observational plane by adopting $E(B-V) = 0.06$ and $\mu=20.77$ \citep{oakes+2022}. The models were computed with a $\Delta t$ of $1\,$Gyr and metallicities following the age-metallicity relation derived by \cite{rusakov+2021}. We note that, in the HST-based star formation history derived by \cite{rusakov+2021}, the metallicity range spanned by young stars covers the metallicity derived for Fornax~6. The CMD in Fig.~\ref{fig: CMD_1Rh} displays a cluster-like morphology. Specifically, we observe a tight and well-populated RGB, and a well-defined and populous SGB/MSTO, albeit with some broadening. The overall morphology is well described by the isochrone with $t\sim3\,$Gyr and $\rm [M/H] = -0.7\,$dex\footnote{The interchangeable use of the global metallicity [M/H] rather than the iron content [Fe/H] is valid as long as the stellar system is characterised by a solar-scaled $\alpha$-elements abundance \citep{salaris+1993}. This is a reasonable assumption for the metal-rich stars of Fornax \citep{tolstoy+2009}.}. 

We compared the CMD centred on Fornax~6 with the field population, following the procedure described in \citet[see also \citealt{dalessandro+2019}]{cabrera-ziri+2016}. This is based on the on-off statistics presented in \cite{knoetig2014}, which quantifies the probability that a star in the CMD of the stellar cluster is a field star.
This method requires the partition of the cluster and field CMDs in a regular grid and, assuming that counts in each bin follow a Poissonian distribution, compares the number of stars counted in a cell of the cluster CMD with that counted in the same cell of the field CMD. With this approach, we can identify which parts of the cluster CMD are more or less compatible with the field. 

\begin{figure}
    \centering
    \includegraphics[width=1\columnwidth]{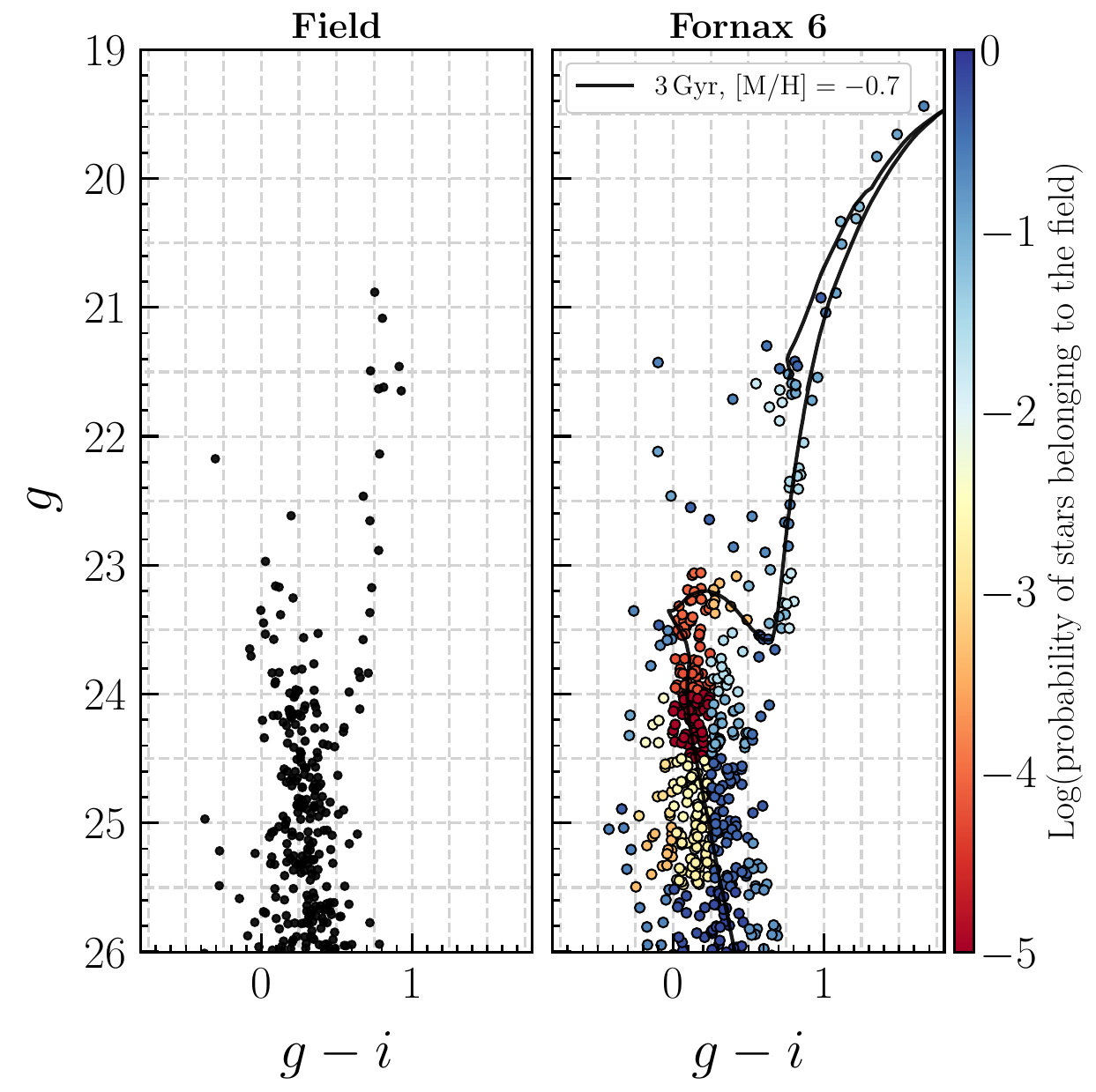}
    \caption{$(g, g-i)$ CMDs of stars within a region of radius $R_h$ centred on random field coordinates (left panel) and Fornax~6 (right panel). Stars in the CMD centred on Fornax~6 are colour-coded according to the logarithm of the probability of being due to the field, with redder colours indicating lower probabilities. A PARSEC isochrone with $t=3\,$Gyr, $\rm[M/H]= -0.7\,$dex, $E(B-V) = 0.06$, and $\mu = 20.77$ \citep{oakes+2022} is also displayed in the right panel. Grey dashed lines show the division of the CMDs into cells (see Sect.~\ref{sec: field_comparison} for details).}
\label{fig: CMD_comparison}
\end{figure}

We again consider the CMD of stars lying within 1$\Rh$ from the centre of Fornax~6. Field stars were selected by considering the same circular area centred on random $(x,y)$ coordinates within the GMOS FOV, avoiding the MUSE footprint (see Fig.~\ref{fig: FOV}). We divided the cluster and field CMDs in a regular grid of $\rm magnitude\times colour = 0.5 \times 0.25\,\rm mag^2$ (see dashed lines in Fig.~\ref{fig: CMD_comparison}), and counted stars falling in each cell. We then applied equation~(23) from \cite{knoetig2014} to compute the probability that the number counted in the cluster CMD cell is due to field stars.
We repeated this experiment several times, selecting different reference fields and cell sizes. 
Figure~\ref{fig: CMD_comparison} shows one of the extractions, where the CMD of the reference field is reported on the left, while the CMD of the overdensity Fornax~6 is on the right. The CMD of Fornax~6 is colour-coded according to the probability calculated for each cell, where blue colours indicate cells with a high probability of being due to the field population, and orange/red colours indicate cells where the number of stars differs significantly from the background.

In all of our draws, stars occupying cells that span the colour-magnitude range $0 < (g-i) < 0.25\,$mag and $23<g<24.5\,$mag (see orange and red points in Fig.~\ref{fig: CMD_comparison}) have a very small probability of being due to the Fornax field.  
If we look at the PARSEC isochrones drawn in Figure~\ref{fig: CMD_1Rh}, these stars fairly correspond to the MSTO and SGB region indicated by the magenta isochrone, which is also displayed in the right panel of Figure~\ref{fig: CMD_comparison} as a black line.
Together with the overall observed CMD morphology, this suggests that, at the location of Fornax~6, there is a significant excess of coeval stars that are $\sim3\,$Gyr old, supporting the hypothesis that the overdensity is indeed a stellar cluster.
However, definitive confirmation that Fornax~6 is such a young cluster requires deep high-resolution observations to mitigate the effects of crowding, blending, and background galaxy contamination, which are likely affecting the accuracy of our photometry at magnitudes fainter than the observed MSTO.

\subsection{On the nature of Fornax~6}
\label{sec: another_cluster?}
Our study aims to examine the true nature of the overdensity Fornax~6 using our new photometric and chemo-kinematical data set. 
The fit presented in Section~\ref{sec: chemo-kinematical_fit} demonstrates the existence of a distinct chemo-kinematical group of stars with respect to the galaxy field population. These stars occupy the locus corresponding to the overdensity as determined via deep photometric data in Section~\ref{sec: structural_properties}, and are characterised by a smaller line-of-sight velocity dispersion than the one measured for field stars. The analysis of the CMD shown in Section~\ref{sec: field_comparison} suggests that the stars tracing the overdensity span a narrow age range, corresponding to a stellar population of $\sim3\,$Gyr. Collectively, these results imply that Fornax~6 is a stellar cluster and place previous conclusions about its nature on firmer ground \citep{wang+2019, pace+2021}.

Yet, if coupled with the observed low luminosity, the measured velocity dispersion ($\sigma_{v,\fnx} = 3.3_{-1.9}^{+1.4}$) of Fornax~6 remains puzzlingly high. Simple stellar population models by \cite{maraston2005} predict $M/L_V = 1.0\,M_{\sun}/L_{\sun}$ for stars with $\met=-0.6\,$dex, $t=3\,$Gyr and a Kroupa initial mass function{\footnote{If we instead consider an age of $t=4\,$Gyr, the same models predict $M/L_V = 1.2\,M_{\sun}/L_{\sun}$}. Consequently, the observed luminosity of Fornax~6 suggests a present-day stellar mass of the order of $M_{\star} = 8.3 \times 10^3\,M_{\sun}$.  Considering virial theorem arguments \citep{portegies_zwart+2010}, the expected line-of-sight velocity dispersion for such a star cluster would be $\lesssim$$1\,\kms$, more than 1.5$\sigma$ below our measurement. 
 
Several factors that cannot be directly quantified with our data set may artificially inflate our measured $\sigma_v$. These include unresolved binaries (see, e.g., \citealt{bradford+2011}), residual field contamination, and underestimated velocity uncertainties \citep{vandeven+2006, kamann+2016}. In addition, the velocity error floor determined by the spectral resolution of MUSE ($1\,\rm km\,s^{-1}$, \citealt{weilbacher+2020}) severely limits the instrument's ability to resolve the velocity dispersion of very low-mass star clusters \citep{zoutendijk+2020}.  On the other hand, the cluster may be unbound. As noted by \cite{pace+2021}, a tidal disruption process can be invoked to explain the high velocity dispersion measured for cluster stars. This scenario would be consistent with the irregular morphology of Fornax~6 that can be seen in Figure~\ref{fig:spatialdistribution}.

The tantalising hypothesis of \cite{penarrubia+2024} that Fornax~6 could be formed by stars temporarily captured within a low-mass ($M \le 10^7\,M_{\sun}$) dark-matter subhalo is also consistent with some of our measurements. These haloes are predicted by the CDM framework, but are challenging to detect because they do not trigger star formation. \cite{penarrubia+2024} argued that in a dSph with properties similar to Fornax, these dark substructures could preferentially trap metal-rich stars due to their cold kinematics (although see \cite{yang+2026} for an alternative view). If correct, this would lead to the emergence of stellar overdensities that are characterised by high mass-to-light ratios and stellar populations identical to those of the surrounding field. To estimate the dynamical mass of Fornax~6, we considered the mass estimator from \cite{errani+2018}. By sampling our posterior distributions of $R_h$ and $\sigma_v$, we obtain $M_{\rm dyn}(<1.8\Rh) = 1.6_{-1.3}^{+2.4} \times 10^5\,M_{\sun}$, which translates into a dynamical mass-to-light ratio $M/L_V(<1.8\Rh) \sim 25$.  As shown in Figure~2 from \cite{penarrubia+2024}, subhaloes with masses below $10^7\,M_{\sun}$ are consistent with both the observed luminosity and the dynamical $M/L_V$ derived for Fornax~6. However, given the CMD morphology of Fornax~6, this hypothesis is disfavoured. Specifically, while the inner regions of the Fornax dSph are dominated by stars of age  $0.2<t<5\,$Gyr \citep{battaglia+2006, delPino+2013,rusakov+2021}, Fornax~6 resembles a single-age population born $\sim3\,$Gyr ago. Nevertheless, higher precision measurements of the velocity dispersion and MSTO region of Fornax~6 will be required to conclusively settle this issue.

\section{Summary and Conclusions}
\label{sec: summary_conclusions}
We combined the exquisite resolution and depth of archival GMOS-S imaging observations with the statistical power of the integral-field spectrograph MUSE to provide a comprehensive photometric and chemo-kinematical view of the stellar overdensity Fornax~6. This has been recently claimed to be the sixth and the second (in projection) most centrally located stellar cluster of the Fornax dSph galaxy (see \citealt{wang+2019,pace+2021}). 
To rule out the possibility that Fornax~6 is, in fact, a field overdensity, we derived a deep CMD that reaches 4 magnitudes below the MSTO, and a large chemo-kinematical catalogue including measurements of $\vlos$ and $\met$ for 131 stars with $S/N\geq15$. These catalogues represent a notable improvement in statistics and photometric accuracy compared to previous ones, allowing a more robust determination of the properties of this putative cluster and an assessment of its true nature.

Overall, the analysis of our new data set supports the hypothesis that Fornax~6 is indeed a stellar cluster. Specifically, a clear overdensity of stars as faint as $g=26\,$mag is visible in the photometric catalogue, although there are signs of irregularity at large radius. By fitting the positions of stars with a flattened Plummer model, we derived a large half-light radius ($\Rh = 8.8^{+1.0}_{-1.3}\,$pc) and an unresolved ellipticity ($e<0.26$, 68\% upper limit). 
Nonetheless, the CMD of stars within $\Rh$ seems to exhibit a well-defined and populated MSTO, SGB and RGB. The CMD morphology is strongly indicative of a cluster-like stellar population with an age of approximately $3\,$Gyr, as determined through the comparison with PARSEC isochrones. The measured total luminosity is $M_V = -5.0 \pm 0.4\,$mag, indicating a low present-day stellar mass. 
Our mixture model applied to the chemo-kinematical data set assigns 43 members to Fornax~6. This number represents more than a twofold increase over the existing spectroscopic sample \citep{pace+2021}. The cluster member stars appear concentrated within the measured $\Rh$, and are characterised by a line-of-sight velocity of $50.8_{-0.7}^{+0.8}\,\kms$ and intrinsic velocity dispersion of $3.3_{-1.9}^{+1.4}\,\kms$. The comparison with the kinematics of field stars ($\overline{v_{los}}_{,\BG} = 56.6_{-1.3}^{+1.4}\,\kms$ and $\sigma_{v,\BG} = 11.4_{-0.9}^{+1.0}\,\kms$) clearly shows that Fornax~6 is a distinct kinematical population.
The corresponding metallicity is $ \met = -0.63_{-0.03}^{+0.03}\,$dex. This is much higher than the other GCs of Fornax, which are characterised by metallicities ranging from $\met =-2.4$ to $\met = -1.4\,$dex \citep{larsen+2022}. According to the HST-based SFH derived by \cite{rusakov+2021}, the kinematically-cold metal-rich population in the inner regions of Fornax formed in multiple bursts at look-back times of $\sim$2.8, 1.8, 0.9 and 0.5 Gyr. The age and metallicity of Fornax~6 are therefore completely consistent with it being a product of this star formation.

Although we have revised downward the previous estimate of the velocity dispersion of Fornax~6 \citep{pace+2021}, our measurement is still higher than what is expected for a star cluster with a stellar mass of a few $10^3\,M_{\sun}$. Several factors could explain this discrepancy, such as unresolved binaries, underestimated velocity uncertainties, and field interlopers. On the other hand, a large measured $\sigma_v$ may indicate that the cluster is unbound or undergoing disruption. Such a scenario could also explain the irregular morphology observed for Fornax~6. 

Alternatively, the size, luminosity and kinematics of Fornax~6 could be consistent with the scenario recently suggested by \cite{penarrubia+2024}, in which small stellar overdensities in dwarf galaxies arise from field-star capture by a parsec-scale dark-matter halo. While this is a tantalising prospect, it is hard to reconcile with the Fornax~6 CMD, which is broadly consistent with a single stellar population. Looking ahead, the {\it Euclid} and {\it Nancy Grace Roman} missions will survey the entire extents of many 
Local Group dwarf galaxies with unprecedented photometric depth and spatial resolution \citep{mellier25,2026arXiv260312981S}.  These data sets will likely reveal many more candidate star clusters and stellar overdensities in these systems \citep[e.g.,][]{2026A&A...706A.185H}, allowing further tests of this intriguing scenario.

\section*{Acknowledgements}
We thank the referee, Andrew Pace, for the helpful comments that improved the paper.
We warmly thank Jorge Pe{\~n}arrubia and Anna Lisa Varri for insightful discussions.
CC and AMNF are supported by the UK Science and Technology Facilities Council [grant number ST/Y001281/1].  AMNF also acknowledges support from UK Research and Innovation (UKRI) under the UK government’s Horizon Europe funding guarantee [grant number EP/Z534353/1]. RP acknowledges the support for this study by the INAF Mini Grant 2025 (Ob.Fu.1.05.24.07.05, CUP C33C24001390005). SKA gratefully acknowledges funding from UKRI through a Future Leaders Fellowship (grants MR/T022868/1, MR/Y034147/1). SM acknowledges funding from the CNES postdoctoral fellowship programme. This study was supported by JSPS KAKENHI Grant Number JP25K07361.
Based on observations collected at the European Southern Observatory under the program 0102.B-0911, and at the Gemini South Observatory under the program GS-2007B-Q-23.

%%%%%%%%%%%%%%%%%%%%%%%%%%%%%%%%%%%%%%%%%%%%%%%%%%
\section*{Data Availability}
The photometric and spectroscopic catalogues will be made available via anonymous FTP at CDS. The description of the catalogues can be found in Appendix~\ref{app: catalogues}. The raw MUSE data are available via the ESO archive, while the GMOS-S images are available in the Gemini Observatory Archive. 
%The inclusion of a Data Availability Statement is a requirement for articles published in MNRAS. Data Availability Statements provide a standardised format for readers to understand the availability of data underlying the research results described in the article. The statement may refer to original data generated in the course of the study or to third-party data analysed in the article. The statement should describe and provide means of access, where possible, by linking to the data or providing the required accession numbers for the relevant databases or DOIs.
%%%%%%%%%%%%%%%%%%%%%%%%%%%%%%%%%%%%%%%%%%%%%%%%%%

\bibliographystyle{mnras}
\bibliography{mnras_biblio} % if your bibtex file is called example.bib

\appendix

\section{The fitting procedure}
\label{subsec:method}

Here, we outline the methodology adopted to investigate the nature of the apparent stellar overdensity Fornax~6. Using the two datasets introduced in the main text - the GMOS photometric catalogue (Section~\ref{sec: photometric_catalogue}) and the MUSE spectroscopic catalogue (Section~\ref{sec: spectroscopy}) - we aim to provide a comprehensive structural and kinematic characterisation of the system. We first describe the general approach used to extract structural and/or chemo-dynamical properties, followed by a detailed explanation of the Bayesian framework employed for the model–data comparison. This methodology has been applied to the different data sets in Sections~\ref{sec: structural_properties} and~\ref{sec: chemo-kinematical_fit}, each providing complementary constraints on the properties of the cluster.

%   intro (inserted in the main text)
%To characterise Fornax~6, we employ a parametric classification scheme based on a mixture modelling approach, in which stars are probabilistically assigned to different components that are determined by maximising the likelihood of a given set of data. The data may comprise the stellar positions on the plane of the sky and/or the stellar kinematics and/or chemistry. The method fits individual stars in a full Bayesian framework, which is key since it allows to estimate the full posterior on the models parameters, enabling a proper computation of confidence intervals, covariances, and avoiding any arbitrary binning of the data. 

%   to the method
\subsection{The method}
Consider a set of stars in a region $\Area$ on the plane of sky. We assume that the probability, $\PP$, that any of these stars is extracted from $\Area$ can be expressed as the sum of $\ncp$ components
\begin{equation}\label{for:gmodel}
    \PP(\DD|\,\btheta) = \sum_{i=1}^{\ncp}\wi\PP_i(\DD|\,\btheta).
\end{equation}
In the above equation, $\DD$ denotes the set of observables within the dataset; $\btheta$ represents a vector of parameters that defines the properties of each component of the model; $\PP_i$ is the probability of the $i$-th component, which is assumed to be normalised to unity, while $\wi$ are weights that determine the relative contributions between components. We require 
\begin{equation}\label{for:wi}
    \sum_{i=1}^{\ncp}\wi = 1,
\end{equation}
to ensure the proper normalisation of $\PP$.

%   our case
In our specific case, we assume that the overall probability distribution is given by the sum of two contributions, which we designate as Fornax 6 ($\PPfnx$) and a background term ($\PPBG$), with $\wii$ setting the relative contribution between the two. Therefore, 
\begin{equation}\label{for:tmodel}
    \PP(\DD|\,\btheta) = \wii\PPfnx(\DD|\,\btheta) + (1 - \wii)\PPBG(\DD|\,\btheta).
\end{equation}

%   strucutral-chemo-dynamical components
In our most general application, the data set consists of the position of each star on the plane of the sky, its line-of-sight velocity and associated uncertainty, as well as its metallicity and corresponding measurement error. Formally, we define the data set $\DD$ as a collection of entries $\{\xj, \yj, \vlosj, \metj\}$, with $j = 1, \dots, N$, where $N$ is the total number of stars. We express the
probability of a star to belong to the $i$-th component as the product
of three distinct terms
\begin{equation}\begin{split}\label{for:pfnx}
    \PP_i(x,y,\vlos, & \met) = \\ &\PP_{i,\SSS}(x,y)\PP_{i,\VV}(\vlos)\PP_{i,\MM}(\met).
\end{split}\end{equation}
In the above equation, $\PP_{i,\SSS}$ accounts for the spatial distribution on the plane of the sky of the $i$-th component, $\PP_{i,\VV}$ for its line-of-sight velocity distribution and $\PP_{i,\MM}$ for its metallicity distribution, with $i=\fnx,\BG$. We point out that, in the above equation, the uncertainties in velocity and metallicity, $\Delta\vlosj$ and $\Delta\metj$ respectively, are not explicit components of the probability function $\PP_i$, even if they are a part of the data set. Instead, their effect is incorporated by convolving $\PP_i$ with an appropriate error distribution, as described in Section~\ref{subsec:bayes}.

\subsubsection{Spatial distribution}
\label{subsec:spatial}

We express the term accounting for the structure of the $i$-th component as
\begin{equation}\label{for:pi}
    \PP_i(x,y) = \frac{S(x,y) \Sigmai(x,y)}{\int_\Area S(x,y)\Sigmai(x,y) \dd x \dd y},
\end{equation}
i.e. the product between a spatial distribution model, $\Sigmai$, and the selection function $S$, which defines the probability that a star is included in the sample. In our case, $S$ depends solely on the position of the stars and serves to delimit the spatial region $\Area$. The denominator in equation~\ref{for:pi} ensures a proper normalisation of $\PP_i$. As in previous works \citep{wang+2019,pace+2021}, we represent the spatial distribution of the Fornax~6 cluster, $\SigmaFnx$, with the flattened \cite{Plummer1911} model
\begin{equation}\label{for:plummer}
    \SigmaFnx(x,y) = \frac{\Rh^2}{\pi (1-e)} \frac{1}{(\Rh^2+m^2)^2},
\end{equation}
with $m$ the elliptical radius
\begin{equation}\begin{split}\label{for:m}
    m^2 \equiv & \biggl[\frac{(x-\xc)\cos\phi - (y-\yc)\sin\phi)}{(1-e)}\biggr]^2 + \\ & [(x-\xc)\sin\phi + (y-\yc)\cos\phi]^2.
\end{split}\end{equation}
In equations \ref{for:plummer} and \ref{for:m}, $e \equiv1- b/a$ represents the ellipticity of the isodensity contours, with $a$ and $b$ being the semi-major and semi-minor axes, respectively; $\Rh$ is the elliptical half-mass radius, i.e. the elliptical radius that contains half of the total projected mass if equation \ref{for:plummer} is normalized to the total mass of the system; $\phi$ is the position angle of the cluster, from north to east. We note that, in equation~\ref{for:m}, $m^2 = \frac{(x-\xc)^2}{q^2} + (y-\yc)^2$ when $\phi=0\deg$, meaning that the semi-major axis is aligned with the position angle axis. Finally,
\begin{equation}
\left\{
\begin{aligned}
    & x = -(\alpha - \alphac)\cos\deltac \\
    & y = \delta -\deltac, 
\end{aligned}
\right.
\end{equation}
where $\xc$ and $\yc$ are the central coordinates of the Fornax~6 cluster with respect to the literature $(\alphac,\deltac)$ from \citet[see their Table~1]{wang+2019}, and $(\alpha,\delta)$ are the Right Ascension and Declination of a star on the plane of the sky. The selection function is
\begin{equation}
S(x,y) =
\left\{
\begin{aligned}
     & 1 \quad {\rm if } \quad (x,y) \in \Area, \\
     & 0 \quad \text{otherwise.}
    \end{aligned}\right.
\end{equation}
We assume that the background is constant in the relevant region, thus
\begin{equation}
    \PPBG = \dfrac{1}{\int_\Area\dd x\dd y}, 
\end{equation}
while it is 0 outside.
In our analysis, stars are selected within a region centred on Fornax~6, the geometry of which depends on the data set. For the GMOS catalogue, the selection function corresponds to a circular region of radius $\Rmax$, whereas for the MUSE data it follows the rectangular footprint of the spectroscopic field of view. The total area considered in the GMOS and MUSE applications is $4.65\,\rm arcmin^2$ and $1.3\,\rm arcmin^2$, respectively.

\subsubsection{Velocity and metallicity distributions}

We assume that both velocity and metallicity distributions of each component follow a Gaussian profile, with the velocity distributions given by
\begin{equation}
    \PP_{i,\VV}(\vlos) = \frac{1}{\sqrt{2\pi}\sigma_{i,\VV}}\exp\biggl[-\frac{(\vlos-\VV_i)^2}{2\sigma^2_{i,\VV}}\biggr],
\end{equation}
and the metallicity distributions by
\begin{equation}
    \PP_{i,\MM}(\met) = \frac{1}{\sqrt{2\pi}\sigma_{i,\MM}}\exp\biggl[-\frac{(\met-\MM_i)^2}{2\sigma^2_{i,\MM}}\biggr].
\end{equation}
In the above equations, $\VV_i$ and $\sigma_{\VV,i}$ are the mean line-of-sight velocity and dispersion of the $i$-th component velocity distribution, while $\MM_i$ and $\sigma_{\MM,i}$ are the mean metallicity and dispersion of the $i$-th metallicity distribution. The total number of free parameters is 14, namely
\begin{equation}\begin{aligned}
    \btheta = \{& \xc, \yc, \Rh, e, \phi, \\ & \VV_\fnx,\sigma_{\VV,\fnx},\MM_\fnx,\sigma_{\MM,\fnx},\\ & \VV_\BG,\sigma_{\VV,\BG},\MM_\BG,\sigma_{\MM,\BG}, \wii
    \}.
\end{aligned}\end{equation}

\subsection{Bayesian fit}
\label{subsec:bayes}

The posterior distribution of the model's parameters $\btheta$ is estimated within a full Bayesian framework of model-data comparison. The likelihood of the model, given the data set $\DD$, is written as
\begin{equation}\label{for:like}
    \LL(\btheta,\DD) = \prod_{j=1}^N(\PP\ast\EE)(\{\DD\}_j|\btheta),
\end{equation}
where $\{\DD\}_j$ is the full set of observables of the $j$-th star in the dataset, including the uncertainties. Thus, $\DD\equiv\{\xj, \yj, \vlosj,\Delta\vlosj, \metj, \Delta\metj\}$. In equation~\ref{for:like}, $\ast$ represents the convolution with the error function $\EE$, which is, in our case, the product of Gaussians with null mean and dispersion equal to $\Delta\vlosj$ and $\Delta\metj$. 

We adopt uniform priors on the model's free parameters. Therefore, according to Bayes' theorem, the posterior distribution is proportional to the product of the likelihood function (equation~\ref{for:like}) and the uniform priors. We sample from the posterior distribution using a Markov Chain Monte Carlo (MCMC) method. Specifically, we use a combination of the differential evolution proposal by \cite{Nelson2014} and the snooker proposal by \cite{terBraak2008}, as implemented in the \texttt{emcee} software library \citep{foreman-mackey+2013}. 

%   further details
For each MCMC run, we initialise a number of walkers equal to six times the number of free parameters. Each walker is evolved for 7000 steps. After visually evaluating model convergence, we always discard an initial burn-in of $~\simeq1000-2000$  steps, depending on the fit, and apply a thinning that is of the order of the autocorrelation length of the chains, $\simeq10-30$. All remaining steps are used to evaluate the posterior distribution of the model's free parameters. Median parameters and any other models' derived quantities are estimated as the 50th percentile of the relevant distributions; the confidence intervals at the $1\sigma$ level are determined by calculating the 16th and 84th percentiles of the distributions, while the $3\sigma$ intervals are computed using the 0.15th and 99.85th percentiles.

In this context, we define the probability of the $j$-th star to belong the $i$-th component as the median value of the distribution of the quantity ($i=\fnx,\BG$)
\begin{equation}\label{for:member}
    \Lprob_j =  \frac{\wi(\PP_i\ast\EE)(\{\DD\}_j|\,\btheta)}{(\PP\ast\EE)(\{\DD\}_j|\,\btheta)}.
\end{equation}
These medians are subsequently rescaled such that, for each star, the total probability $\Lprob_\fnx + \Lprob_\BG  = 1$. 

\section{Posteriors}
In this Appendix, we present the cornerplots illustrating the marginalised one- and two-dimensional posterior distributions from the fitting procedure (Appendix~\ref{subsec:method}). Specifically, Fig.~\ref{fig:cornerplot_MUSE} shows the result of the chemo-kinematical-spatial fit performed on the MUSE data, while Fig.~\ref{fig:cornerplot_GMOS} shows the result of the spatial-only fit performed on the GMOS data. 

\label{app:appA}
\begin{figure*}
    \centering    \includegraphics[width=0.9\vsize,angle=90]{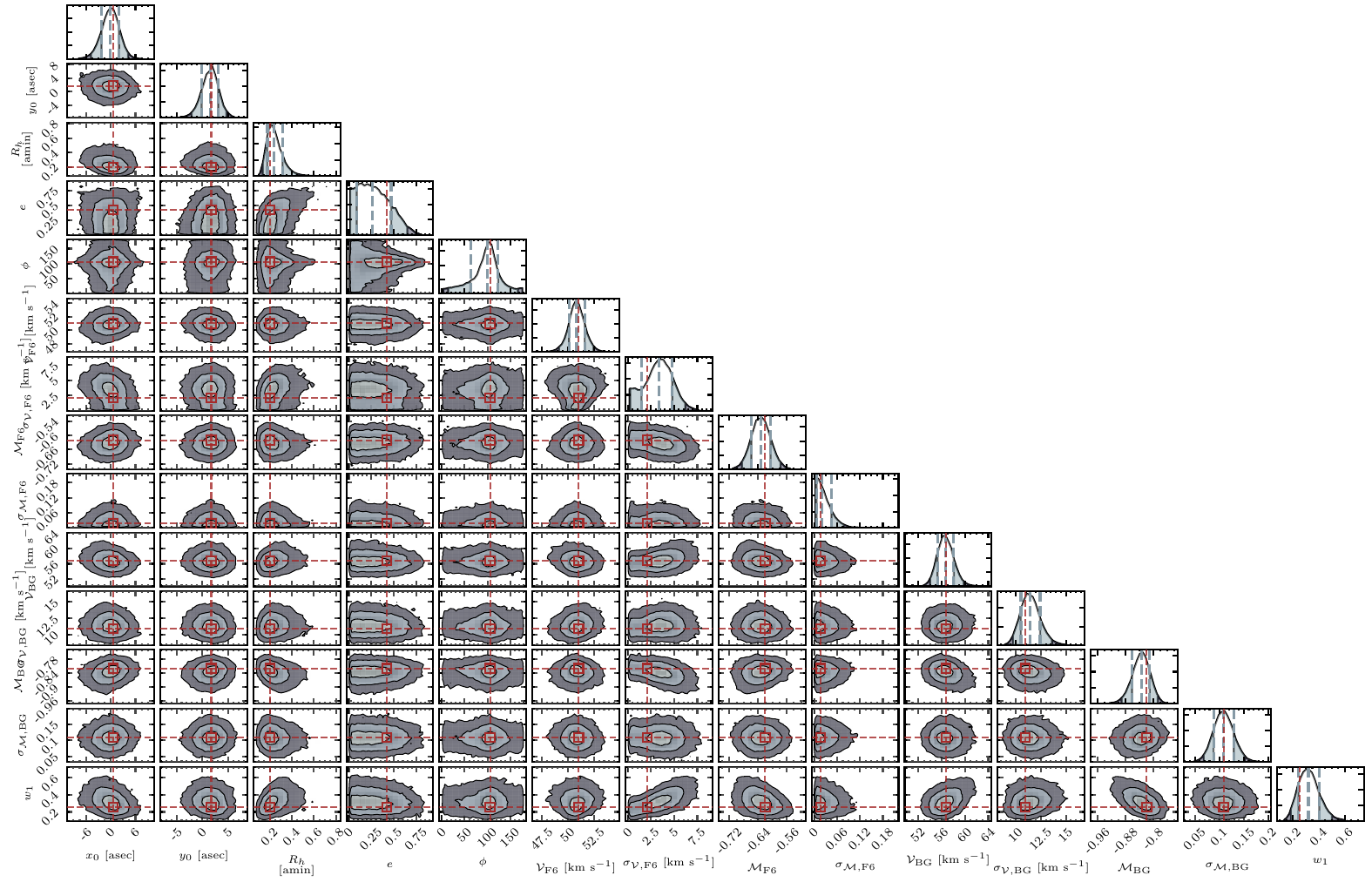}
    \caption{Marginalised one- and two-dimensional posterior distributions on the MUSE model parameters. Different areas within the two-dimensional marginalised distributions represent regions that enclose 68\%, 95\%, and 99\% of the total probability, respectively. The dashed grey vertical lines in the one-dimensional marginalised distributions correspond to the 16th, 50th and 84th percentiles, used to estimate the uncertainties over the model's free parameters. The vertical dashed red lines in the marginalised one-dimensional distributions, and the squares in the marginalised two-dimensional distributions show the position of the model with the highest posterior among the ones explored.}\label{fig:cornerplot_MUSE}
\end{figure*}

\begin{figure*}
    \centering
    \includegraphics[width=1\textwidth]{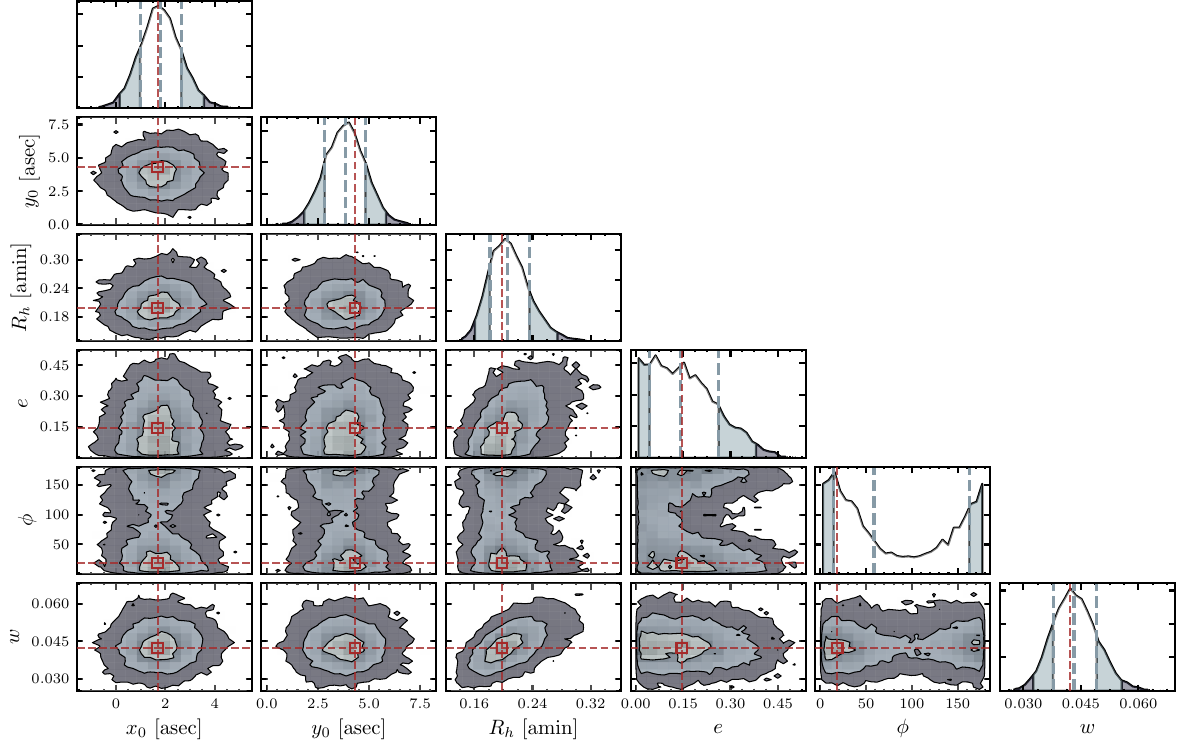}
    \caption{Same as \ref{fig:cornerplot_MUSE}, but for the GMOS model parameters.}\label{fig:cornerplot_GMOS}
    %Marginalised one- and two-dimensional posterior distributions on the GMOS model parameters. Different areas within the two-dimensional marginalised distributions represent regions that enclose 68\%, 95\%, and 99\% of the total probability, respectively. The dashed grey vertical lines in the one-dimensional marginalised distributions correspond to the 16th, 50th and 84th percentiles, used to estimate the uncertainties over the model's free parameters. The vertical dashed red lines in the marginalised one-dimensional distributions, and the squares in the marginalised two-dimensional distributions show the position of the model with the highest posterior among the ones explored.}\label{fig:cornerplot_GMOS}
\end{figure*}

\section{Catalogues description}
\label{app: catalogues}
In this Appendix, we describe the content of the catalogues that we made public in the CDS via anonymous ftp.
\subsection{GMOS-S photometric catalogue}
The GMOS-S photometric catalogue was derived as described in Sect.~\ref{sec: reduction_GMOS}. We publish a version of the catalogue that includes all sources detected in our photometric analysis. However, we remind the reader that our structural fit was performed by considering sources within a circular region of radius $73\arcsec$ ($1000\,$px) from the centre of Fornax~6. The reference system for the instrumental coordinates is the co-added $i$-band image. Table~\ref{tab: catalogue_GMOS} provides a column-by-column description of the file.

To select highly probable star sources, we recommend the following selection:
\begin{itemize}
    \item \texttt{f\_chi == 1} and \texttt{f\_sh == 1}, which select sources satisfying our selection in $chi$ and $sharpness$ as described in Sect.~\ref{sec: photometric_catalogue};
    \item \texttt{f\_col == 1}, which selects sources satisfying the colour cuts applied in Sect.~\ref{sec: photometric_catalogue} to exclude background sources and spurious objects.
\end{itemize}

\begin{table}
%  \centering
%    \renewcommand{\arraystretch}{1.75}
\caption{Description of the columns in the catalogue based on GMOS-S data.}
    \begin{tabular}{ll}
        \hline
        Column & Description  \\
        \hline
        ID   &  Source identifier  \\
        $X$  &  $X$-position in reference image (px) \\
        $Y$  &  $Y$-position in reference image (px) \\
        R.A. & Right Ascension (deg)   \\
        Dec  & Declination (deg)  \\
        g\_mag & AB magnitude in the $g$ filter (mag) \\
        g\_err   & DAOPHOT photometric uncertainty in the $g$ filter (mag) \\
        i\_mag & AB magnitude in the $i$ filter (mag) \\
        i\_err   & DAOPHOT photometric uncertainty in the $g$ filter (mag)  \\
        chi  & DAOPHOT $chi$ for global PSF fit \\
        sh   & DAOPHOT $sharpness$ for global PSF fit \\
        f\_chi & Flag for the $chi$ parameter \\
        f\_sh  & Flag for the $sharpess$ parameter \\
        f\_col & Flag for colour selection \\
        \bottomrule
\end{tabular}
\label{tab: catalogue_GMOS}
\end{table}

\subsection{MUSE chemo-kinematical catalogues}
We publish two catalogues with measurements of line-of-sight velocity and metallicity, derived as described in Sect.~\ref{sec: spectroscopy}. The first catalogue contains $v_{los}$ and $\met$ measurements for all star sources (i.e., background galaxies were excluded; see Sect.~\ref{fig: background_galaxies}) found in the MUSE datacube with $S/N \geq 10$. Stars' positions and magnitudes were derived via crossmatch with the GMOS-S catalogue (see previous paragraph). Table~\ref{tab: catalogue_MUSE_TOT} provides a column-by-column description of the file.
To work with sources for which reliable measurements are available, we recommend applying the following cuts:
\begin{itemize}
    \item \texttt{SNR $\ge$ 15};
    \item \texttt{[Fe/H]\_err $< 0.3$}; 
    \item \texttt{(g-i) $> 0.5$};
    \item exclude the source with \texttt{ID\_MUSE} == 2119, which does not satisfy the colour selection applied to the GMOS-S photometric catalogue.
\end{itemize}

\begin{table}
\begin{minipage}{0.48\textwidth}
%  \centering
%    \renewcommand{\arraystretch}{1.75}
\caption{Description of the columns in the catalogue based on MUSE data. Magnitudes and positions are derived from PSF fitting of GMOS-S imagery.}
    \begin{tabular}{ll}
        \hline
        Column & Description  \\
        \hline
        ID\_MUSE   &  Source identifier  \\
        R.A. & Right Ascension (deg)   \\
        Dec  & Declination (deg)  \\
        g\_mag & AB magnitude in the $g$ filter (mag) \\
        g\_err   & DAOPHOT photometric uncertainty in the $g$ filter (mag) \\
        i\_mag & AB magnitude in the $i$ filter (mag) \\
        i\_err   & DAOPHOT photometric uncertainty in the $g$ filter (mag)  \\
        SNR  & PampelMuse signal-to-noise  \\
        LOS\_V   & Spexxy best-fit line-of-sight velocity value ($\kms$) \\
        LOS\_V\_err & Spexxy error on line-of-sight velocity\footnote{The reported uncertainty was obtained by adding in quadrature $1\,\kms$ (see Sect.~\ref{sec: spectroscopy}).} ($\kms$) \\
        Teff  & Spexxy best-fit effective temperature value (K)\\
        Teff\_err & Spexxy error on effective temperature\footnote{\label{note11}Output uncertainties were increased following the prescription of \cite{husser+2016}.} (K) \\
%        Log\_g & Spexxy best-fit surface gravity value \\
        $\met$  & Spexxy best-fit metallicity value (dex) \\
        $\met$\_err & Spexxy error on metallicity\footref{note11} (dex)\\
        \bottomrule
\end{tabular}
\label{tab: catalogue_MUSE_TOT}
\end{minipage}
\end{table}

The second catalogue contains only the sources used in the fitting procedure. In addition to the flags mentioned above, we excluded five stars with very low metallicities ($\met < -1.4\,$ dex). This file provides the probability of belonging to either the Fornax~6 cluster or the Fornax dSph field, following equation\,\ref{for:member}. We also report stars' coordinates used during the fit, defined as offsets from the literature Fornax~6's centre (see Appendix\,\ref{subsec:spatial}). Table~\ref{tab: catalogue_MUSE_FIT} provides a column-by-column description of the file.

\begin{table}
\begin{minipage}{1\textwidth}
%  \centering
%    \renewcommand{\arraystretch}{1.75}
\caption{Description of the columns in the catalogue used in the fit of MUSE data. Magnitudes and positions are derived from PSF fitting of GMOS-S imagery.}
    \begin{tabular}{ll}
        \hline
        Column & Description  \\
        \hline
        ID\_MUSE   &  Source identifier  \\
        R.A. & Right Ascension (deg)   \\
        Dec  & Declination (deg)  \\
        Delta\_X & R.A. coordinates on the plane tangent to the sky at the literature centre of Fornax~6\footnote{\label{note0}From \cite{wang+2019}.}  (deg)   \\
        Delta\_Y  & Dec coordinates on the plane tangent to the sky at the literature centre of Fornax~6\footref{note0} (deg)  \\
        g\_mag & AB magnitude in the $g$ filter (mag) \\
        g\_err   & DAOPHOT photometric uncertainty in the $g$ filter (mag) \\
        i\_mag & AB magnitude in the $i$ filter (mag) \\
        i\_err   & DAOPHOT photometric uncertainty in the $g$ filter (mag)  \\
        SNR  & PampelMuse signal-to-noise  \\
        LOS\_V   & Spexxy best-fit line-of-sight velocity value ($\kms$) \\
        LOS\_V\_err & Spexxy error on line-of-sight velocity\footnote{The reported uncertainty was obtained by adding in quadrature $1\,\kms$ (see Sect.~\ref{sec: spectroscopy}).} ($\kms$) \\
        $\met$  & Spexxy best-fit metallicity value (dex) \\
        $\met$\_err & Spexxy error on metallicity\footnote{\label{note1}Output uncertainties were increased following the prescription of \cite{husser+2016}.} (dex) \\
        Prob\_F6  & Probability of belonging to the Fornax~6 cluster \\
        Prob\_Fnx & Probability of belonging to the Fornax dSph field \\
        \bottomrule
\end{tabular}
\label{tab: catalogue_MUSE_FIT}
\end{minipage}
\end{table}

\end{document}